\documentclass{article}

\usepackage{arxiv}

\usepackage[utf8]{inputenc} 
\usepackage[T1]{fontenc}    
\usepackage{hyperref}       
\usepackage{url}            
\usepackage{booktabs}       
\usepackage{amsfonts}       
\usepackage{nicefrac}       
\usepackage{microtype}      

\usepackage{mathptmx}      
\usepackage{courier}       
\usepackage{type1cm}      
\usepackage{amsmath}
\usepackage{amsthm}
\usepackage{amssymb,bbm}
\usepackage{xcolor}
\usepackage{adjustbox}
\usepackage{url}

\usepackage{makeidx}        
\usepackage{graphicx}        
                             
\usepackage{multicol}        
\usepackage{natbib}
\usepackage[bottom]{footmisc}
\usepackage{bbm}
\usepackage{relsize}
\usepackage{float}

\newtheorem{thm}{Theorem}
\newtheorem{proposition}[thm]{Proposition}
\newtheorem{corollary}[thm]{Corollary}

\title{Modal clustering of matrix-variate data}

\author{
  Federico Ferraccioli\\
  Dipartimento di Scienze Statistiche \\
    Università degli Studi di Padova \\
       Via Cesare Battisti, 241, 35121 Padova PD\\
  \texttt{federico.ferraccioli@unipd.it} \\
   \And
 Giovanna Menardi \\
  Dipartimento di Scienze Statistiche \\
       Università degli Studi di Padova \\
       Via Cesare Battisti, 241, 35121 Padova PD\\
  \texttt{menardi@stat.unipd.it} \\
}

\begin{document}

\maketitle

\begin{abstract}
    The nonparametric formulation of density-based clustering, known as modal clustering, draws a correspondence between groups and the attraction domains of the modes of the density function underlying the data. Its probabilistic foundation allows for a natural, yet not trivial, generalization of the approach to the matrix-valued setting, increasingly widespread, for example, in longitudinal and multivariate spatio-temporal studies. 
    In this work we introduce nonparametric estimators of matrix-variate distributions based on kernel methods, and analyze their asymptotic properties. Additionally, we propose a generalization of the mean-shift procedure for the identification of the modes of the estimated density. Given the intrinsic high dimensionality of matrix-variate data, we discuss some locally adaptive solutions to handle the problem. We test the procedure via extensive simulations, also with respect to some competitors, and illustrate its performance through two high-dimensional real data applications. 
\end{abstract}


\section{Introduction}
\label{sec:1}

The analysis of complex data in the form of matrices represents an active area of research. Classical examples are longitudinal studies and multivariate spatio-temporal data, where statistical observations are represented by vectors of variables, measured on subjects over different times or locations, or modern analyses where the matrix structure is intrinsic to the problem, as in the case of image data, adjacency matrices representing networks, covariance or similarity structures etc. 
While the focus is usually intended to the former case, collections of data of such type are often referred to as \emph{three-way}, with two ways associated to the row and column dimension of each matrix-variate observation and the third one represented by subjects. For an extensive review, see \citet{kroonenberg08}.  

A vast body of literature has focused on the development of supervised methods for matrix-valued data. Under this umbrella, we shall include regression models with multivariate observations gathered over time \citep[see, e.g. ][for a standard account]{diggle2002} or penalized linear regression models with matrix-valued response \citep{zhou2014regularized}. Specific classes of matrix-valued data, such as semi-definite positive matrices \cite[see for example][]{dryden2009non} and orthogonal matrices \cite[see for example][]{chakraborty2019statistics}, have also received recent attention.   
Far less attention has been devoted to the unsupervised case. Historical contributions here are the Tucker3 and the Candecomp/Parafac models \citep[][Ch 5.7 and, respectively, 4.6]{kroonenberg08} and developments, as well as various attempts to reduce data dimension via principal component analysis or akin methods \citep[see, e.g., ][Ch. 1 and 3]{sakata16}. 

For the aim of clustering, longitudinal or functional data methods have been largely proposed for grouping subjects based on a single feature measured over time, while only few contributions refer to the observation of three-way structures. Stemming from \citet{basford1985}, \citet{viroli2011} develops model-based clustering building on mixtures of matrix-variate Normal distributions and its bayesian counterpart \citep{viroli2011model}. Similarly, \citet{gallaugher2018} propose skewed matrix-variate distributions for unsupervised and semi-supervised classification. 

With a somewhat different aim in mind, further scattered examples which are worth to mention are \citet{wang2019}, which perform tensor decomposition to cluster individuals and tissues in gene expression data; \citet{vermunt07}, where a hierarchical mixture model is used to group longitudinal matrix-variate data in clusters which possibly vary across times, and \citet{vichi_etal07} who perform clustering of subjects and factorial dimensionality reduction of variables and occasions of three-way data. 

In this work we extend the \emph{nonparametric} formulation of density-based clustering, also known as \emph{modal} clustering, to the framework of matrix-variate data. Here, clusters are identified as the ``domains of attraction'' of the modes of the true density underlying the data \citep{stuetzle2003}. 
The inherent notion of cluster is hence not linked to any predefined shape, and determining the number of clusters is an integral part of the estimation procedure. 
While with a different rationale, nonparametric clustering shares with its more widespread parametric counterpart a sound probabilistic foundation, which also allows for a precisely defined population goal. In fact, the issue of density estimation, usually addressed via nonparametric methods, assumes a key role in order to approximate the ideal population goal of modal clustering, along with the operational search of the modal regions.  

After providing an overview on the modal clustering approach, we introduce a kernel estimator for matrix-variate density functions. We then study its asymptotic properties, also with reference to the problem of optimal bandwidth selection.
Due to the intrinsic high dimensionality of matrix-variate data, which impacts on both the accuracy of the estimate and the computational complexity, we explore some local solutions to handle the problem. 
Additionally, we propose an extension of the mean-shift procedure for the identification of the modes of the estimated density.  Finally we perform an extensive simulation study and illustrate the performance of the proposed method on two sets of real data.

\section{An overview on modal clustering}

In the following, and throughout the paper, we will denote by lower-case symbols both scalar and vector-valued objects, whereas matrices will be denoted by uppercase letters. With some abuse, we will also use the same notation to indicate random objects and associated realizations, and specify explicitly their nature when it is not clear from the context.   
In a standard multivariate setting, \emph{modal} clustering relies on the assumption that the observed data $\mathcal{X} = (x_1, \ldots, x_N)$ are realizations of a multidimensional random variable $x \in \mathbb{R}^P$ with (unknown) probability density function $f$. The modes of $f$ are regarded to as representatives of the clusters, which are in turn represented by their domains of attraction. Broadly speaking, if the underlying density is ﬁgured as a mountainous landscape, and modes are its peaks, clusters are the ‘regions that would be ﬂooded by a fountain emanating from a peak of the mountain range’ \citep{chacon2015}. Morse theory allows a more formal framing of the problem, by defining  
clusters as the stable manifolds of the gradient flow associated with the local maxima of $f$. These are represented by the sets of all the points which converge to the same mode by following the gradient ascent paths of the true density. 

While the population clustering goal is defined precisely in terms of features of the underlying density, this is in practice unknown, and needs to be estimated. The issue is far from being trivial, as the estimated density determines the modal regions, and hence governs the final clustering.
A standard choice, within the class of nonparametric methods, is the kernel density estimator
\begin{equation}\label{eq:kde_standard}
	\hat{f}(x; h) = \frac{1}{Nh^P}\sum_{n = 1}^{N}K\left(h^{-1}(x - x_{n})\right),
\end{equation}
where the kernel $K$ is a probability density on $\mathbb{R}^P,$ symmetric around zero, and the bandwidth $h > 0$ is a scale parameter defining the degree of smoothing. 
While the choice of the kernel is known not to have a strong impact on the performance of $\hat f$, a proper selection of the bandwidth turns out to be crucial. Small values of $h$ lead to an undersmoothed density estimate, with the possible appearance of spurious modes, whereas too large values result in an oversmoothed density estimate, possibly hiding relevant features.

A further aspect to account for in modal clustering is to operationally characterize the clusters as the domains of attraction of the density modes. 
Most of the contributions in this direction take their steps from the \emph{mean-shift} algorithm \citep{fukunaga1975} which, starting from a generic point $y^{(0)}$, recursively shifts it uphill to a local weighted mean of the data, along the direction
of the gradient of its kernel estimate: 
\begin{equation*}
    y^{(s + 1)} = \sum_{n = 1}^{N} w_{n, h}(y^{(s)})x_{n}. 
    \label{eq:ms_step}
\end{equation*}
The weights $w_{n,h}(\cdot)$ are specified as normalized components of the gradient of the kernel function. Hence, the mean shift is shown to be a gradient ascent algorithm based on a normalized kernel estimator of the gradient. The convergence of the sequence $\{y_0, y_1, \ldots, y_s, \ldots\}$ to a local mode of \eqref{eq:kde_standard} has been studied under various assumptions by  \citet{ghassabeh2015sufficient} and \citet{arias2016estimation}.


A partition of the data is finally obtained by grouping in the same cluster the observations ascending to the same mode of the density. 
The reader may refer to \citet{menardi2016} and references therein for insights on modal clustering. 

\section{Matrix-variate extension of modal clustering}
\subsection{Kernel density estimation of matrix-variate data}

Let $X_{1}, \dots, X_{N}$ be a sample of \emph{i.i.d.}  realizations of a $P\times T$ random matrix $X =\{x_{p,t}\}_{p=1, \ldots, P, t=1, \ldots,T}$, which we shall assume to be defined on the vector space $\mathbb{R}^{P, T}$. The (unknown) distribution of $X$ is naturally described by some probability density function $f: \mathbb{R}^{P, T} \mapsto \mathbb{R}_+, $ with $\int_{\mathbb{R}^{P, T}} f(X) \mathrm{d}X = 1, $ being the component-wise integral of $f$ on its support. 

Consider an integrable kernel $K : \mathbb{R}^{P,T} \mapsto \mathbb{R}_{+}$, with unit integral and spherically symmetric, i.e. 
$\int_{\mathbb{R}^{P,T}}X K(X)dX = 0$. We define
the \emph{kernel density estimator} for matrix-variate data as
\begin{equation}\label{eq:kde}
\hat{f}(X; h) = \frac{1}{Nh^{P\cdot T}}\sum_{n = 1}^{N}K(h^{-1}(X - X_{n})), \qquad h > 0.
\end{equation}
With the above established convention on defining matrix-variate integrals as their component-wise counterpart, and the same for derivatives, most of standard results on kernel density estimators extend naturally to the matrix-variate setting. The Mean Integrated Square Error (MISE) admits forthwith the usual representation 
\citep[e.g.][p. 28]{chacon2018} 
    \begin{align}
       \text{MISE} &(\hat{f}(X; h)) = \mathbb{E}\int_{\mathbb{R}^{P,T}}(\hat{f}(X;h) - f(X))^{2}dX \nonumber \\
        & =  \int_{\mathbb{R}^{P,T}}\text{Var}(\hat{f}(X;h))dX + \int_{\mathbb{R}^{P,T}}\text{Bias}^{2}(\hat{f}(X; h))dX \nonumber\\
       &  = \text{IV}(\hat{f}(X;h)) + \text{ISB}(\hat{f}(X;h))\label{eq:mise}
\end{align}
and, likewise, its dependence on the bandwidth is not easily disclosed, as the latter enters implicitly via the integrals involving the kernel. 
To highlight the effect of the bandwidth, it is useful to derive an asymptotic approximation of the MISE. To this aim, we further assume the following:
\begin{itemize}
    \item[(i)] $f$ is square integrable and twice differentiable, with all its second order partial derivatives bounded, continuous and square integrable;
    \item[(ii)] the kernel $K$ is, in turn, square integrable, with finite second order moments $\int X\otimes X K(X)dX = m_{2}(K)\mathrm{vec}\mathbb{I}_{P\times T},$ and $m_{2}(K) = \int x_{p,t}^{2}K(X)dX,$ $p = 1, \dots, P, t= 1, \ldots T.$ The symbol $\otimes$ here denotes the Kronecker product; 
    \item[(iii)]the bandwidths $h = h_{N}$ form a positive sequence, such that $h \rightarrow 0$ and $N^{-1}h \rightarrow 0$ as $N \rightarrow \infty$. 
\end{itemize} 
Then, the following holds.
\begin{proposition}
	The asymptotic mean integrated squared error (AMISE) for $\hat{f}(\cdot; h)$ is 
	\begin{equation*}
	\text{AMISE}(\hat{f}(\cdot; h)) = N^{-1}h^{-(P\cdot T)}R(K) + \frac{1}{4}h^{4}m_{2}(K)^{2}R(\Delta f) ,
	\end{equation*}
	and it is minimized by 
	\begin{equation}\label{eq:h_opt}
	h_{\text{AMISE}} = \left(\frac{(P\cdot T) R(K)}{m_{2}(K)^{2}R(\Delta f)}\right)^{\frac{1}{(P \cdot T)+4}}N^{-\frac{1}{(P\cdot T)+4}},
	\end{equation}
	where, for a square integrable function $a:\mathbb{R}^{P \times T} \mapsto \mathbb{R}$ we denote by $\Delta a = \sum_{p = 1}^{P}\sum_{t = 1}^{T} \frac{\partial^{2}a(X)}{\partial x_{p,t}^{2}}$ the Laplacian operator and $R(a) = \int_{\mathbb{R}^{P,T}} a(X)^{2}dX$ its square integral.
\end{proposition}
\begin{proof}
    See Appendix.
\end{proof}
Hence, as for vector-valued data, the approximately optimal bandwidth converges to zero as $N$ increases at the rate $N^{-\frac{1}{(P\cdot T)+4}}$. Also, 
the optimal solution \eqref{eq:h_opt} relies on the knowledge of the true $f$, and hence cannot be directly used to define the optimal smoothing amount. Consistently with the vector case, automatic bandwidth selection can be built by first estimating either the 
MISE or its asymptotic version (AMISE), and then minimising such estimate to yield a bandwidth computed solely from the data. Standard approaches based on cross-validation, bootstrap, or based on replacing the target density with a given parametric model in the expressions of the MISE/AMISE can be easily extended to the matrix-variate framework.

In fact, as for the standard multivariate settings, the use of a scalar bandwidth $h$ may result in a poor flexibility, and richer classes of parameterizations may be alternatively considered. The maximal extent of flexibility would require the awkward use of a four-way structure whose entries would reflect all the possible covariances between pairs of the $X$ components. 
Alternatively, the vectorization operator may be easier to this aim, by mapping $\mathbb{R}^{P,T}$ to $\mathbb{R}^{P\cdot T}$ and stacking the column vectors of $X$ underneath each other in order from left to right. With this representation, a full, unconstrained bandwidth $H$ takes the form of a symmetric, semidefinite matrix $P\cdot T\times P \cdot T,$ yet with some limitations from the algebraic and computational points of view. 

A remarkable simplification may be induced by certain Kernels, for which a separable structure of $H$ is available, so that an equivalent specification represents the matrix-variate $P\times T $ Kernel as a special case of a $PT-$variate Kernel with bandwidth  ${H} = U \otimes V,$ with $U$ and $V$ symmetric positive definite matrices of dimension $P\times P$ and $T \times T$, respectively. Elliptical models belong to this family, and are defined as
\begin{align*}
K(V^{-1/2}(X-X_n)U^{-1/2}) = |V|^{-\frac{P}{2}}|U|^{-\frac{T}{2}}g\left(-\frac{1}{2}\mathrm{tr}(V^{-1}(X-X_n)^{\top}U^{-1}(X-X_n))\right),
\end{align*}
where $g:\mathbb{R}\mapsto \mathbb{R}_{+}$ is such that $\int_{\mathbb{R}_+} z^{pt-1}g(z^2)dz < \infty$ \citep{caro2016matrix}. The matrices $U$ and $V$ act independently on the rows and columns 
and are easier to handle than a full specification of the matrix $H$, that might result challenging even when $P$ and $T$ are small.  
By following the same steps as in Proposition 1, we may obtain the expression of the AMISE when $H = U \otimes V$. In this case, the first term depends on the determinant $|H|^{-1/2}$ instead of $h^{-(P \times T)}$, while the second term involves the full Hessian instead of the simple Laplacian. The general result, however, does not lead to an explicit formula for the optimal bandwidth matrix.

Note that the simplest case, where $U = h_{U}\mathbb{I}_{P}$ and $V = h_{V}\mathbb{I}_{T}$, reduces to the form 
\begin{align}
\label{eq:simple_k} 
    K((h_{U}h_{V})^{-1}(X - X_n)) = h_{U}^{-P}h_{V}^{-T}g\left(-\frac{1}{2(h_{U}h_{V})^{2}}\mathrm{tr}((X - X_{n})^{\top}(X - X_{n}))\right),\nonumber
\end{align}
hence the choice of two distinct smoothing parameters for rows and columns brings back to the scalar case as an effect of the separability of the scale matrix $H$.

Within the class of elliptical kernels, we may set $g(\cdot) = (2\pi)^{-\frac{P\cdot T}{2}} \exp(\cdot)$ and obtain 
the matrix Normal density, 
a natural candidate for the kernel function which 
plays a pivotal role in the matrix-variate framework, as for the univariate and 
multivariate settings \citep[see, e.g,][]{gupta_nagar18}.





\subsubsection{Adaptive kernel}
As an overall problem shared by nonparametric tools, kernel estimators are known to strongly suffer from the curse of dimensionality. On one side, the required sample size to achieve an acceptable accuracy becomes disproportionately large as the dimensions increases, leading to intractable problems, even computationally. On the other side, 
in high dimensions, the sparsity of data leads much of the probability mass to flow to the tails of the density, possibly averaging away features in the highest density regions and giving rise to the birth of spurious modes.

These arguments could discourage from the application of modal clustering on matrix-variate data, which are intrinsically high-dimensional, except for very small values of $P$ and $T$. In fact, nonparametric estimates can still be useful to coarsely describe the data structure, and often allowing different amounts of smoothing is advisable to capture local structures of the data.   
In this direction, adaptive estimators build on the idea that for data-sparse regions, a large bandwidth is needed to compensate for the few nearby data points, and conversely, for data-dense regions, a small bandwidth applies smoothing in a small region due to the presence of many nearby data points. As a general principle, we may distinguish between \emph{balloon} and \emph{sample point} estimators, which replace $h$ in equation \eqref{eq:kde} by $h(X)$ and $h(X_i)$ respectively. See \citet[][\S 4.1]{chacon2018} for an overview in the multivariate setting.  
Within these classes, we consider, for the matrix-variate setting, a $k$-nearest neighbor ($k$-NN) extension of the two  estimators, defined as  
\begin{eqnarray}\label{eq:knn_b}
    \hat{f}_B(X; k) &=& \frac{1}{N\delta_{k}(X)^{P \cdot T}}\sum_{n = 1}^{N}K\left({\delta_{k}(X)}^{-1}(X - X_{n})\right),\\\label{eq:knn_sp}
    \hat{f}_{SP}(X; k) &=& \frac{1}{N}\sum_{n = 1}^{N}\frac{1}{\delta_{k}(X_n)^{P \cdot T}} K\left(\delta_{k}(X_n)^{-1}(X - X_{n})\right),
\end{eqnarray}
where $\delta_{k}(X) = || X - X^{(k)}||_F$ is the Frobenius distance of $X$ from its $k$-th nearest neighbour $X^{(k)}.$ 
 
\subsection{Mean shift for matrix-variate data}\label{sec:mean-shift}

Once the density has been estimated via the matrix-variate extension discussed so far, clusters can be associated to the domains of attraction of the modes of such density, to be intended as high-density subsets of the sample space surrounding the (matrix-variate) local maxima of the density. 

With this regard, the following proposition states that the hill-climbing property of the mean-shift algorithm still holds in the matrix-variate setting. 
\begin{proposition}
Consider a differentiable kernel $K: \mathbb{R}^{P,T}\mapsto \mathbb{R}_+$,  with  unit  integral, and spherically symmetric. Let $\kappa(\cdot): \mathbb{R}_+ \mapsto \mathbb{R}$ be a function such that $K(X) = \frac{1}{2}\kappa(\text{tr}(X^{\top}X))$ and its derivative $\kappa'(u) \leq 0.$

Then, starting at $Y^{0}\in \mathbb{R}^{P,T},$ the sequence defined by 
\begin{equation}
	Y^{(s+1)} = Y^{(s)} + M(Y^{(s)})= \sum_{n = 1}^{n} w_{n,h}(Y^{(s)}) X_{n}\label{eq:msmatrix_step}
\end{equation}
describes a gradient ascent algorithm on \eqref{eq:kde},
with 
\begin{equation}\label{eq:msmatrix}
M(Y) = \frac{\sum_{n=1}^N\kappa'\left(h^{-2}\mathrm{tr}((X_{n} - Y)^\top(X_{n} - Y))\right)(X_{n} - Y)}{\sum_{n = 1}^{n}\kappa'\left(h^{-2}\mathrm{tr}((X_{n} - Y)^\top(X_{n} - Y))\right)}
\end{equation}
denoting the mean-shift and
\begin{equation*}
    w_{n,h} (Y) =  \frac{\kappa'\left(h^{-2}\mathrm{tr}((X_n - Y)^\top (X_n - Y)) \right)} {\sum_{n = 1}^{N} \kappa'\left(h^{-2}\mathrm{tr}((X_n - Y)^\top (X_n - Y)) \right)}.
\end{equation*}
 	
\end{proposition}
\begin{proof}
    See Appendix.
\end{proof}
While loosing its interpretation as an iterative weighted average of the observations, the gradient ascent nature of the mean-shift may be derived also for more complex structures of the bandwidths. When a kernel function with separable structure $H= U\otimes V$ is used, for instance, the \eqref{eq:msmatrix} becomes
\begin{equation*}
M(Y) =  \frac{\sum \kappa'\left(\text{tr}(V^{-1}(X_{n} - Y)^\top U^{-1}(X_{n} - Y))\right) V^{-1/2}(X_{n} - Y)U^{-1/2}} {\sum \kappa'\left(\text{tr}(V^{-1}(X_{n} - Y)^\top U^{-1}(X_{n} - Y))\right) 
}    
\end{equation*}
\noindent
with some simple mathematical manipulation.

With respect to the adaptive estimator \eqref{eq:knn_sp}, the same 
proposition holds, with the only caution of replacing $h$ with $\delta_k(X_n).$
Conversely, the same arguments do not generally apply to the balloon estimator \eqref{eq:knn_b}, since the kernel depends on $X$ also through $\delta_k(X)$, and therefore does not allow to derive a general expression for its gradient. An exception in the multivariate case occurs when the kernel is chosen among the beta family, where the problem simplifies remarkably \citep{duong2016nearest}. The same naturally extends to the matrix variate case. Specifically, when 
a uniform kernel on the unit $PT$-ball 
is selected, the gradient ascent property of the mean-shift is shown to hold with extreme computational efficiency, as stated by the following result.

\begin{figure*}[t]
    \centering
    \begin{minipage}{0.23\linewidth}
            \includegraphics[width = \linewidth]{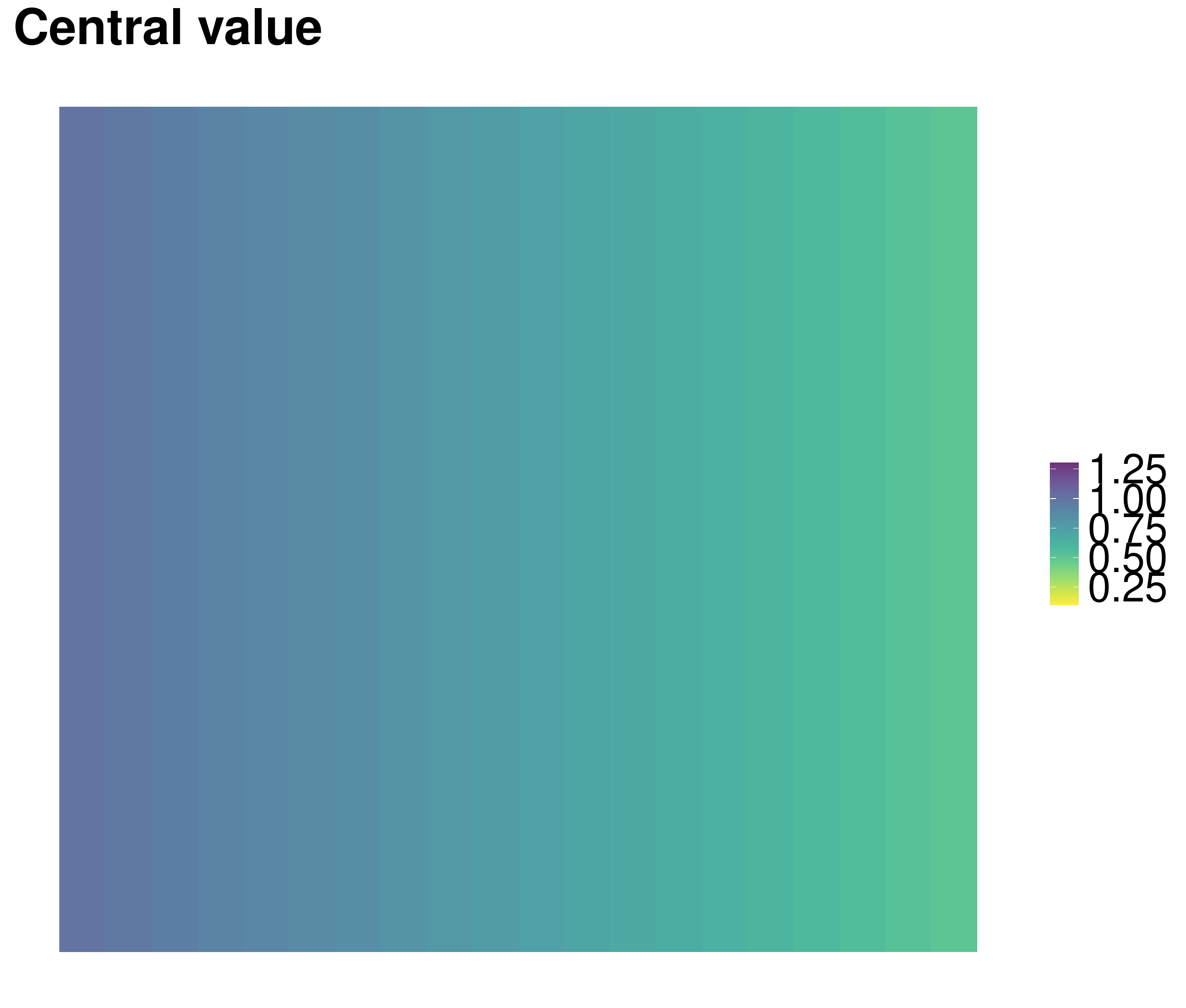}
    \end{minipage}%
    \begin{minipage}{0.23\linewidth}
            \includegraphics[width = \linewidth]{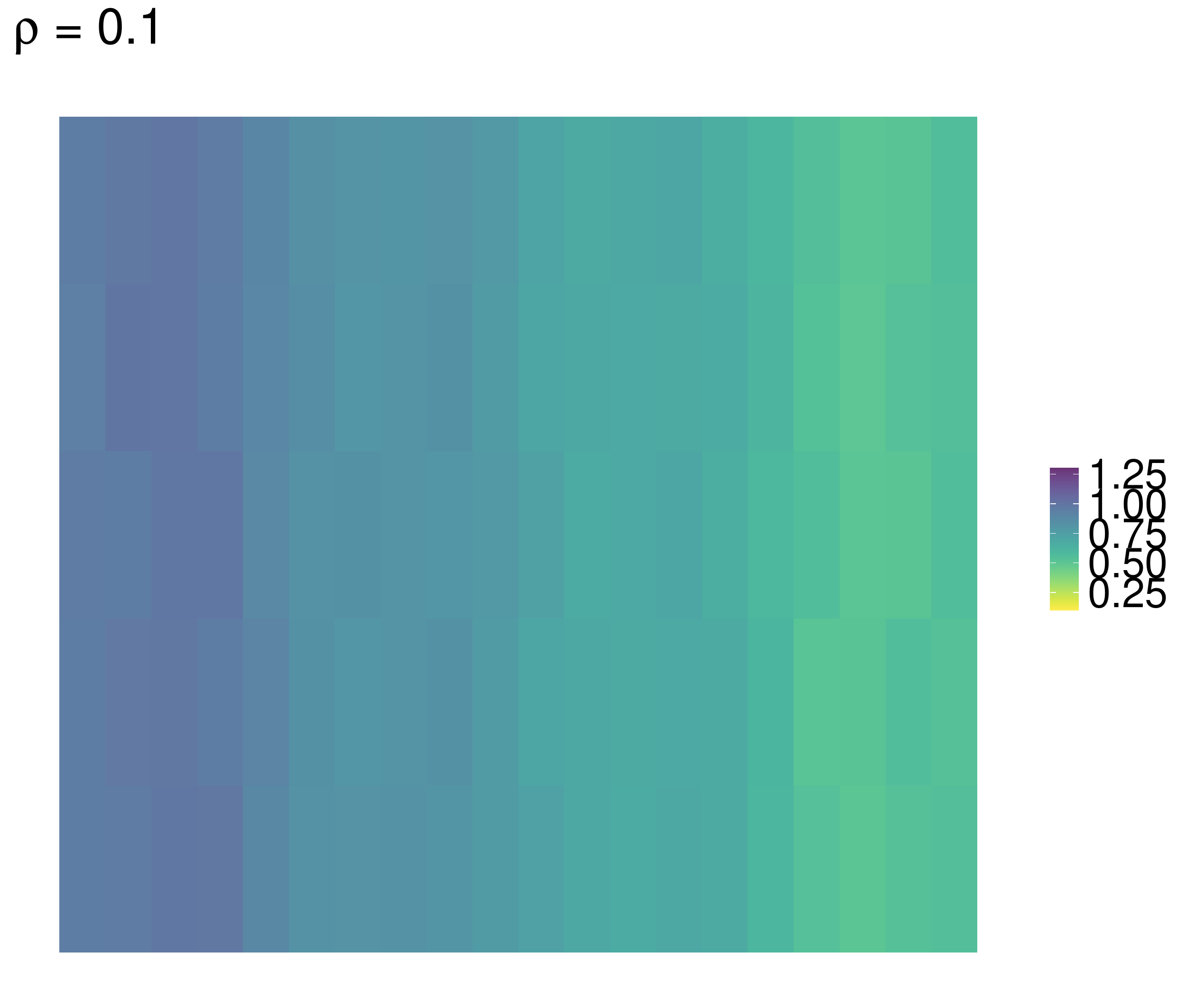}
    \end{minipage}%
        \begin{minipage}{0.23\linewidth}
            \includegraphics[width = \linewidth]{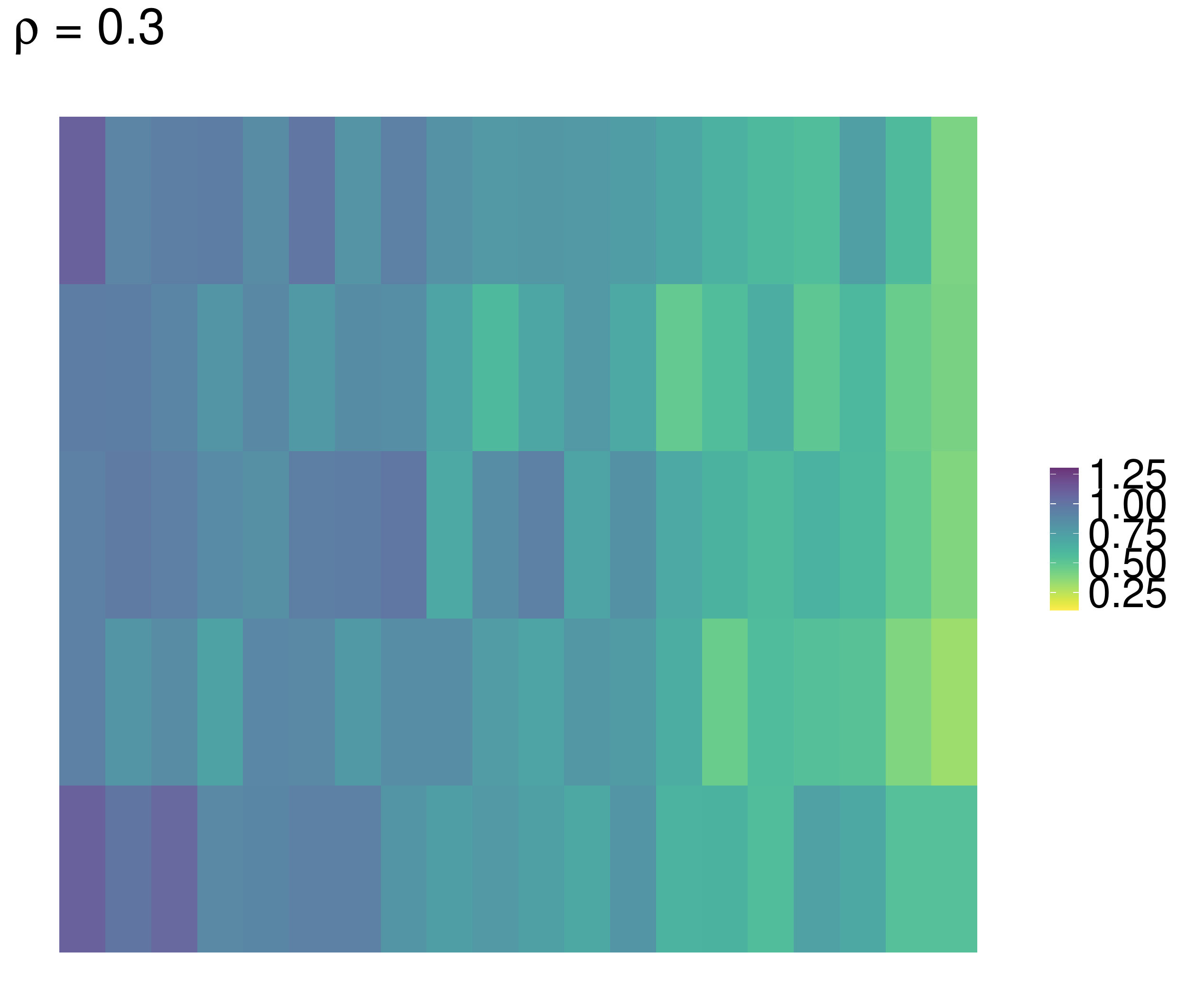}
          \end{minipage}%
      \begin{minipage}{0.23\linewidth}
            \includegraphics[width = \linewidth]{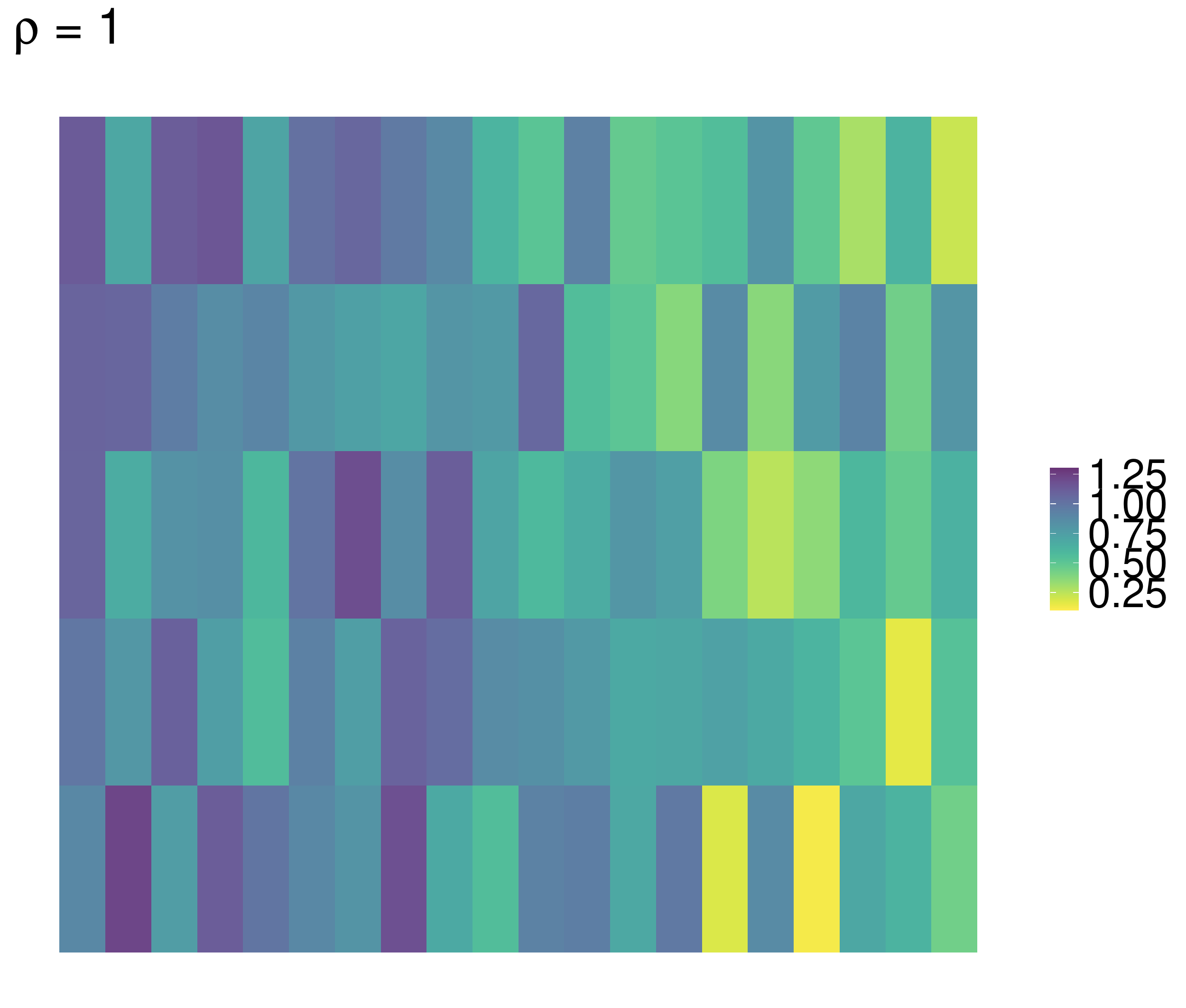}
      \end{minipage}
\caption{Graphical representation of an example of matrix prototype $M$ (left), where color intensities are associated to different values of the matrix entries. The second, third and fourth panel display three random realizations of the random matrix associated to $M,$ for increasing values of $\rho.$}
\label{fig:gen}
\end{figure*}

\begin{corollary}
Consider the adaptive estimator in \eqref{eq:knn_b}, with  $K(X)=\nu_0^{-1}\mathbbm{1}\{X \in B_{PT}(0,1)\}$
and $B_{PT}(0,1)$ the unit ball centered at $0$, with  hypervolume $\nu_0.$ 
Then, the mean shift sequence \eqref{eq:msmatrix_step} takes the form
\begin{equation*}
    Y^{(s+1)} = \frac{1}{k}\sum_{X_n : X_n \in B_{PT}(Y^{(s)},\delta_k(Y^{(s)})}  X_{n}.
\end{equation*}
\end{corollary}
    The proof follows the same steps of the one of Proposition 2. See also \cite{duong2016nearest} for the multivariate case.

\section{Simulations}
\subsection{Settings}

In this Section we present an extensive simulation study with the aim of evaluating the performances of the proposed approach to cluster three-way data, with respect to the following aspects: (1) different group configurations, sample sizes, data dimension; (2) the use of different formulations of kernel-type matrix-variate estimators; (3) comparison with some competitors. 

In the case of matrix-valued data, generating random samples with some interesting, nontrivial structure to be disclosed is awkward, and literature is quite scarce. Some Gaussian matrix-variate examples can be found in \citet{viroli2011} and \citet{viroli2011model}. 
Here we follow a different route, based on multidimensional Discrete Cosine Transform \citep[DCT,][]{strang1999discrete}, a transformation technique for data compression, widely used in digital media and imaging. DCT is
able, in principle, to handle and control for structures with varying degrees of complexity. 
For each cluster, we define a matrix prototype $M$ of size $P \times T$, and express it as 
\begin{equation*}
  M = L^{\top} \Omega R,
\end{equation*}
where $L$ and $R$ are two orthogonal matrices with dimensions $P\times P$ and $T \times T$, respectively, that contain the basis of the decomposition. The matrix $\Omega$, of dimension $P\times T$ is the so called DCT, and its elements $\omega_{p,t}$ are computed stemming from the entries $m_{p,t}$ of $M$ as (see \cite{makhoul1980fast})
\begin{equation*}
 \omega_{p, t} = 4\sum_{i = 1}^{P}\sum_{j = 1}^{T}m_{i, j}\cos\left(\frac{\pi(2i - 1)(p - 1)}{2P}\right)\cos\left(\frac{\pi(2j - 1)(t - 1)}{2T}\right).
    \end{equation*}
The cosine factors are the elements of the matrices $L$ and $R^{\top}$, respectively.
A random matrix $X$ of size $P\times T$ is then built starting from $M$ by replacing each of the entries $\omega_{p,t}$ of $\Omega$ by $\omega_{p,t}+\epsilon u,$ with $\epsilon \sim N(0,\sigma^2)$ and $u \sim Bin(1, \rho)$.  
In practice, a random proportion $\rho$ of DCT coefficients is contaminated with normal error of zero mean and fixed variance. 
The role of $\rho$ is twofold: on one side, it determines, along with $\sigma^2,$ the amount of sample variability and, on the other side, it governs the shape and distribution of the clusters. While setting $\rho$ equal to one determines the generation of matrix-variate spherical normal clusters, any lower proportion leads to some departure from such distribution. Figure \ref{fig:gen} presents a graphical example of matrix prototype, and three random realizations associated to increasing values of $\rho.$  

\begin{figure*}[t]
      \centering
      \begin{minipage}{0.3\linewidth}
            \includegraphics[width = \linewidth]{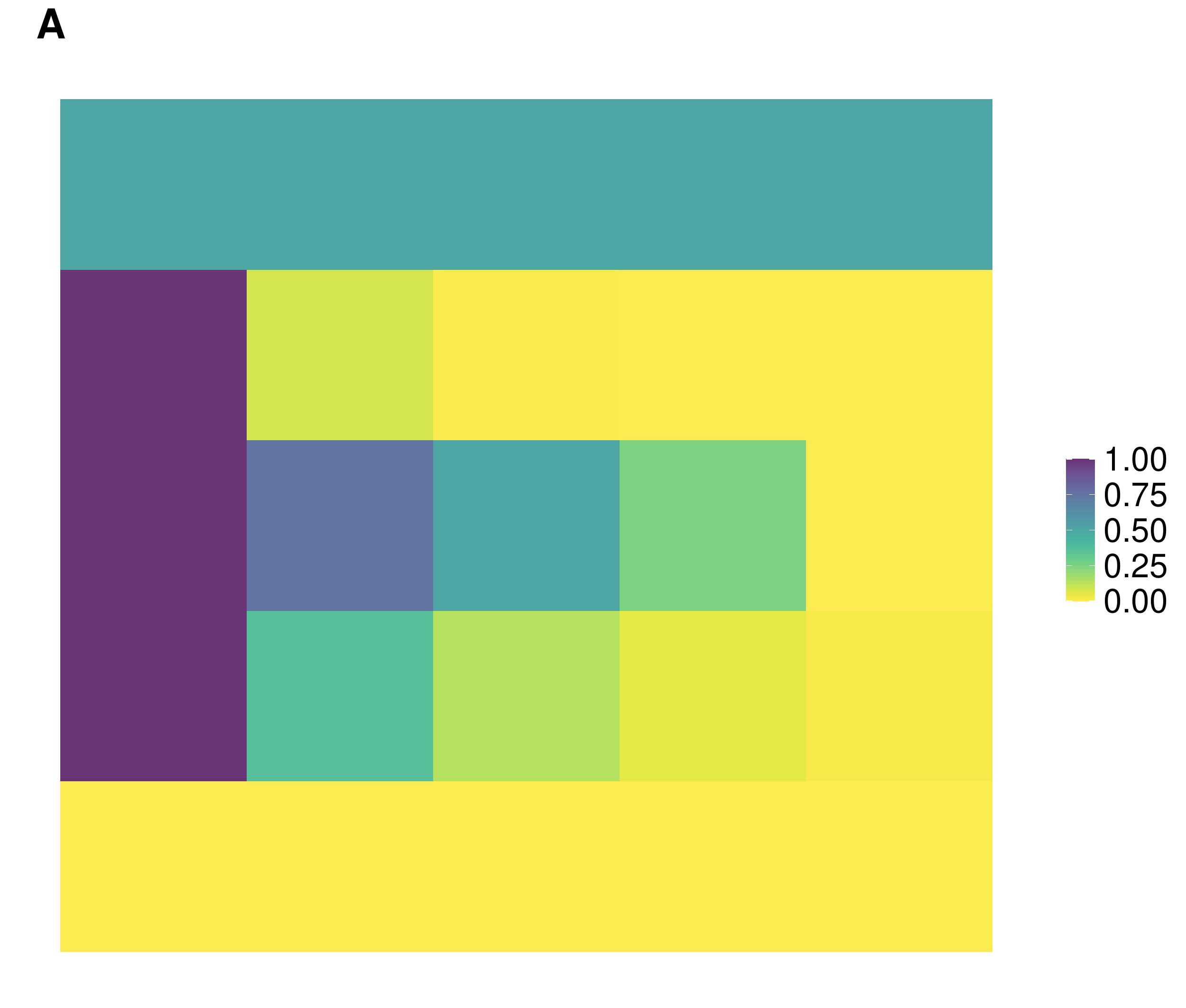}
      \end{minipage}%
        \begin{minipage}{0.3\linewidth}
            \includegraphics[width = \linewidth]{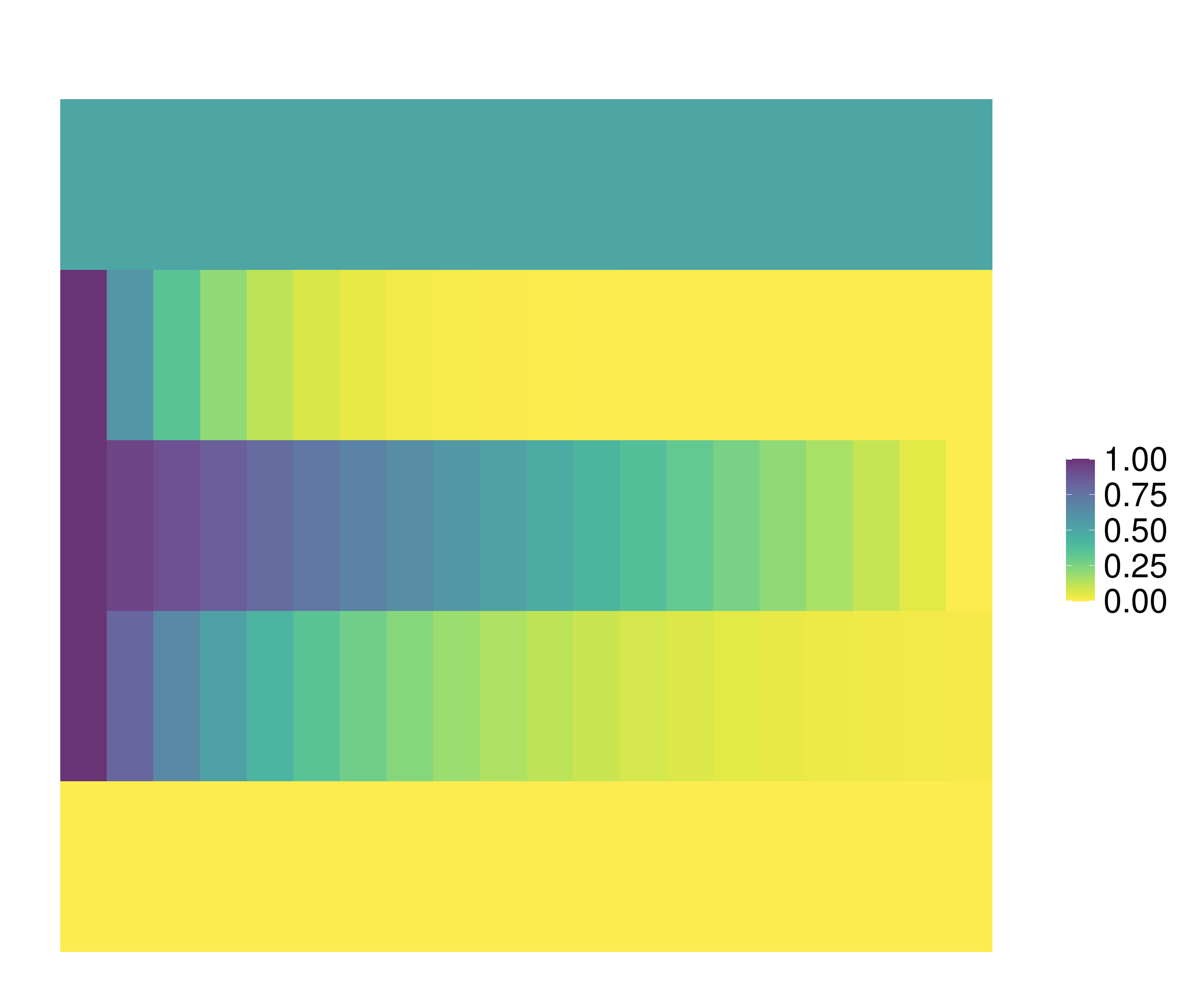}
          \end{minipage}
          \begin{minipage}{0.3\linewidth}
            \includegraphics[width = 0.95\linewidth]{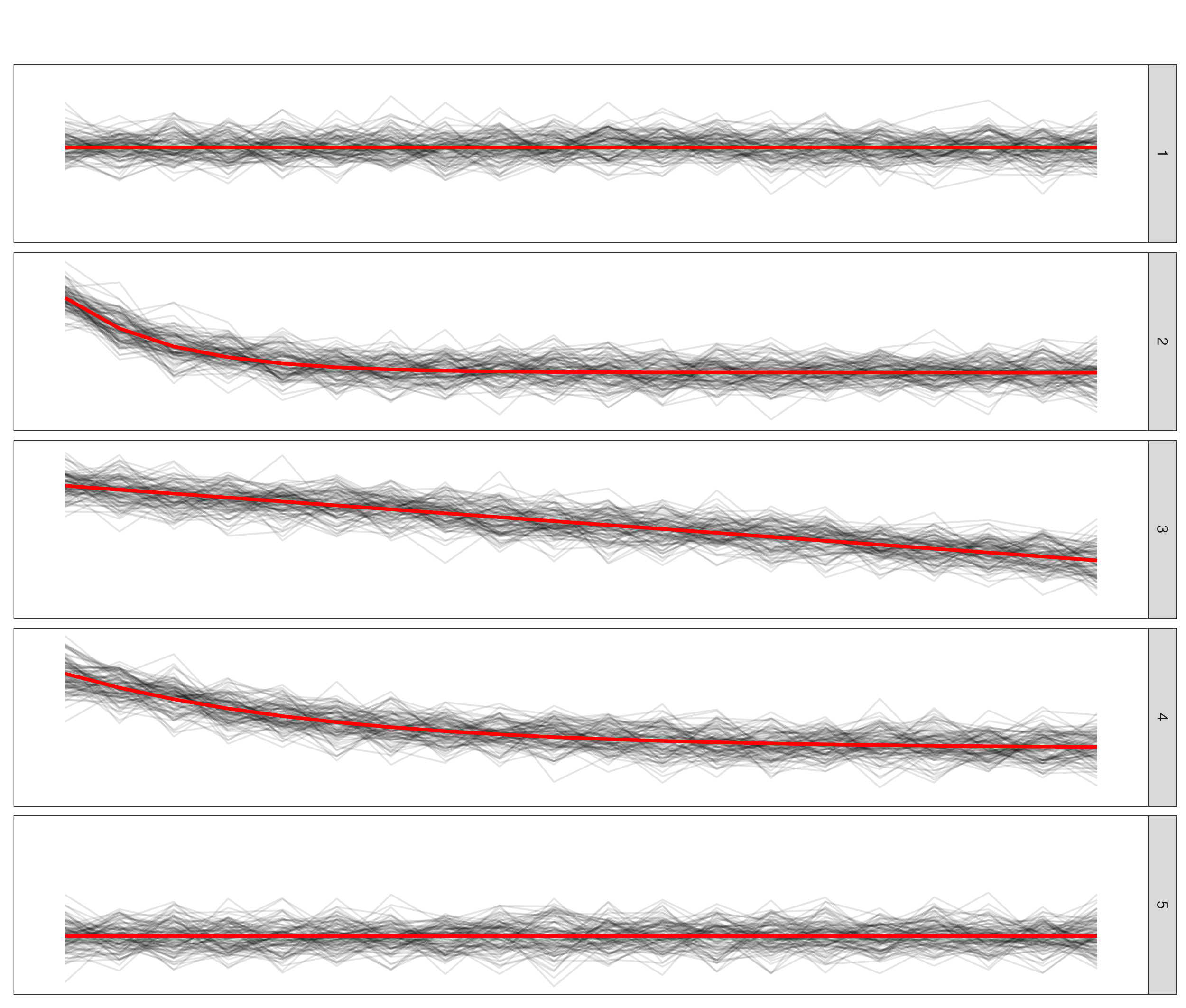}
      \end{minipage}%
    \\
      \begin{minipage}{0.3\linewidth}
            \includegraphics[width = \linewidth]{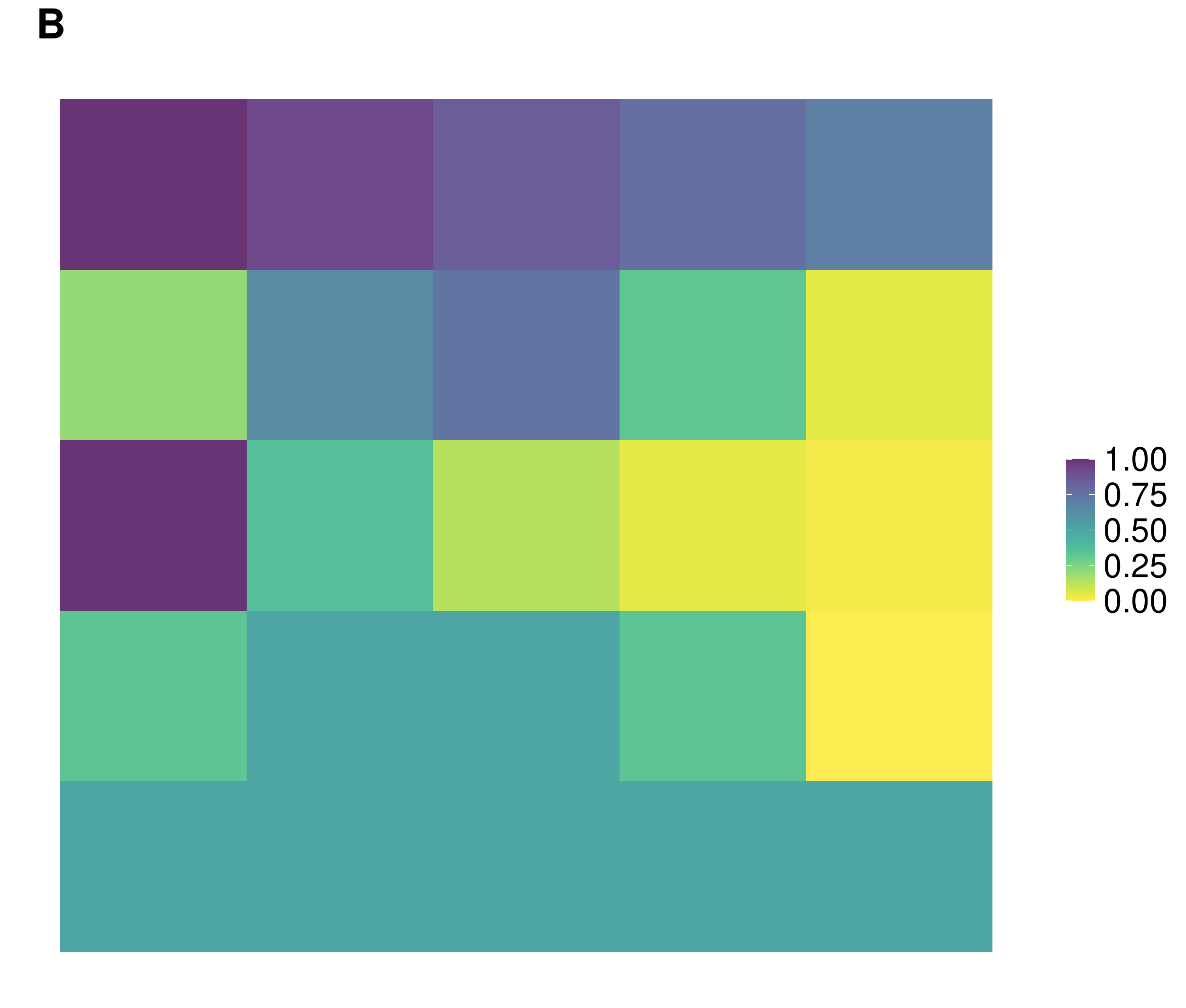}
      \end{minipage}%
        \begin{minipage}{0.3\linewidth}
            \includegraphics[width = \linewidth]{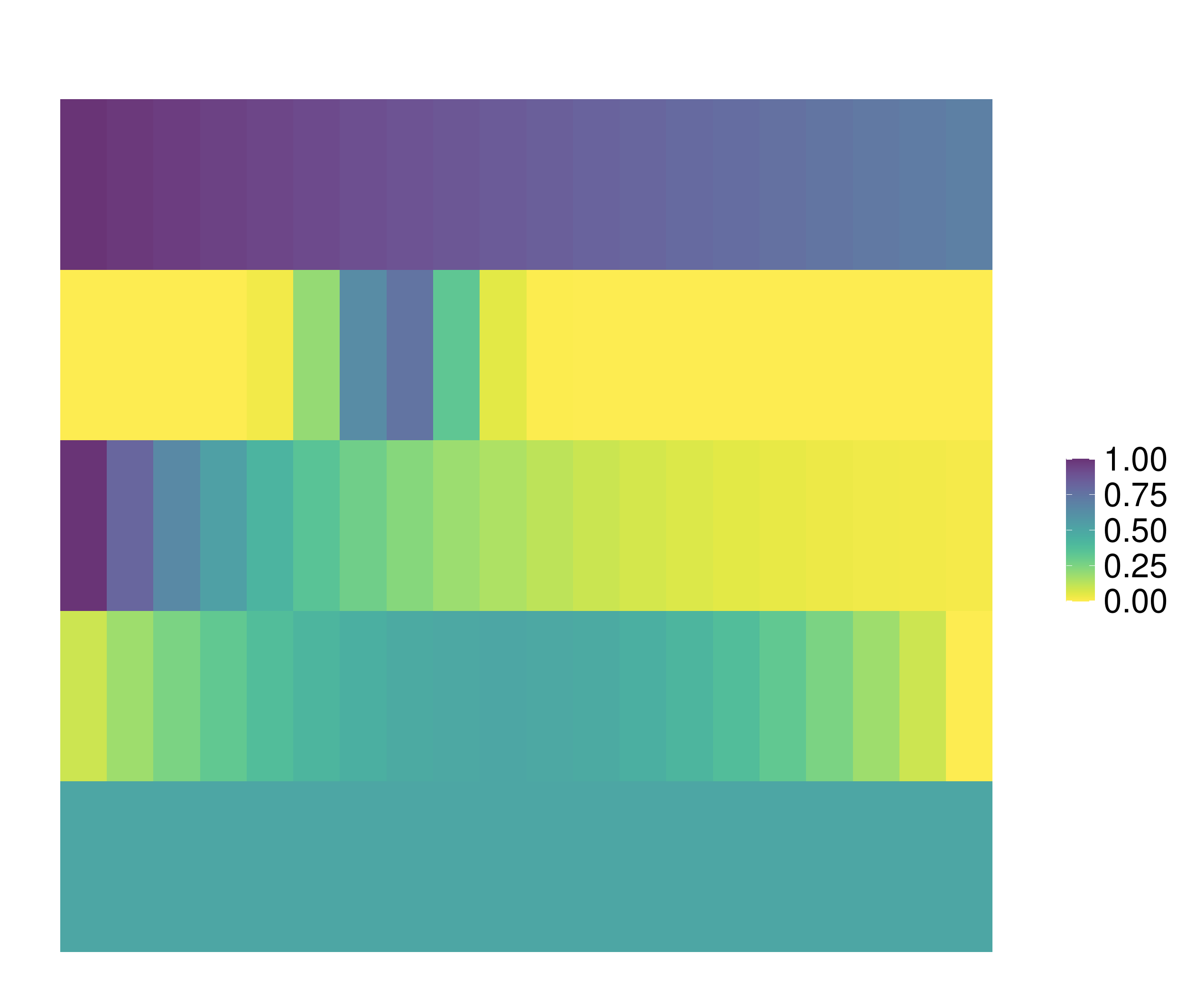}
          \end{minipage}
            \begin{minipage}{0.3\linewidth}
            \includegraphics[width = 0.95\linewidth]{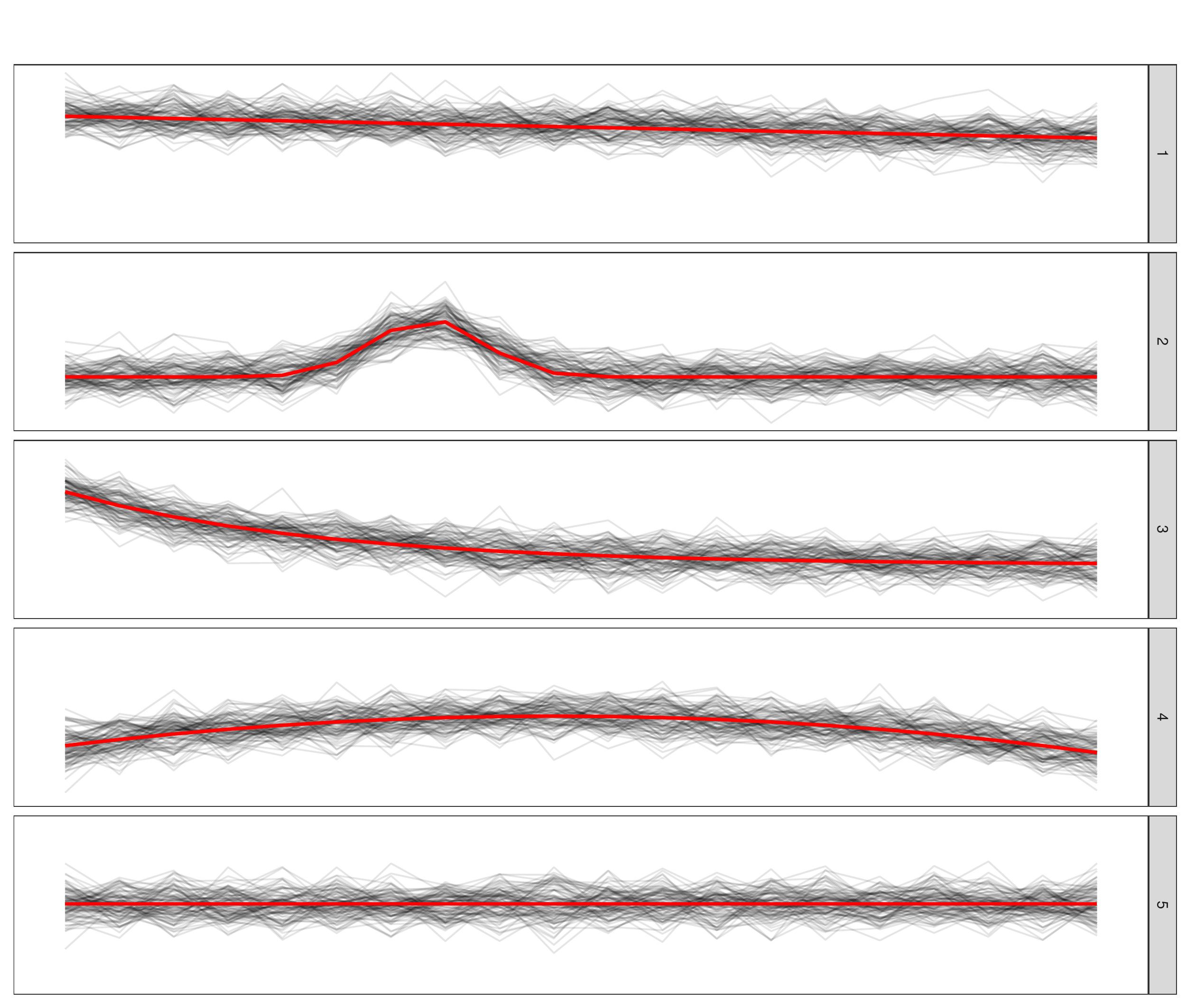}
      \end{minipage}\\
       \begin{minipage}{0.3\linewidth}
            \includegraphics[width = \linewidth]{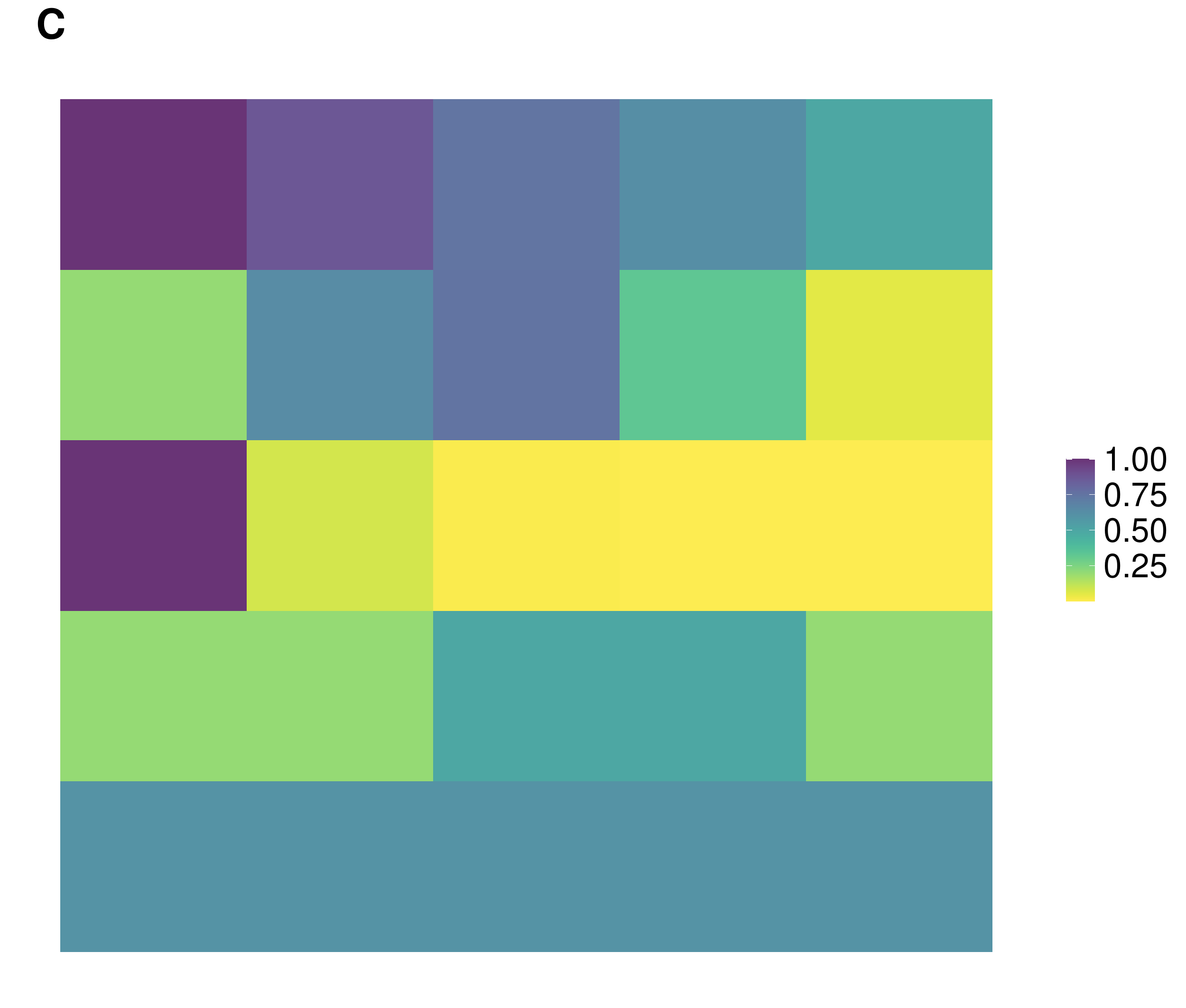}
      \end{minipage}%
        \begin{minipage}{0.3\linewidth}
            \includegraphics[width = \linewidth]{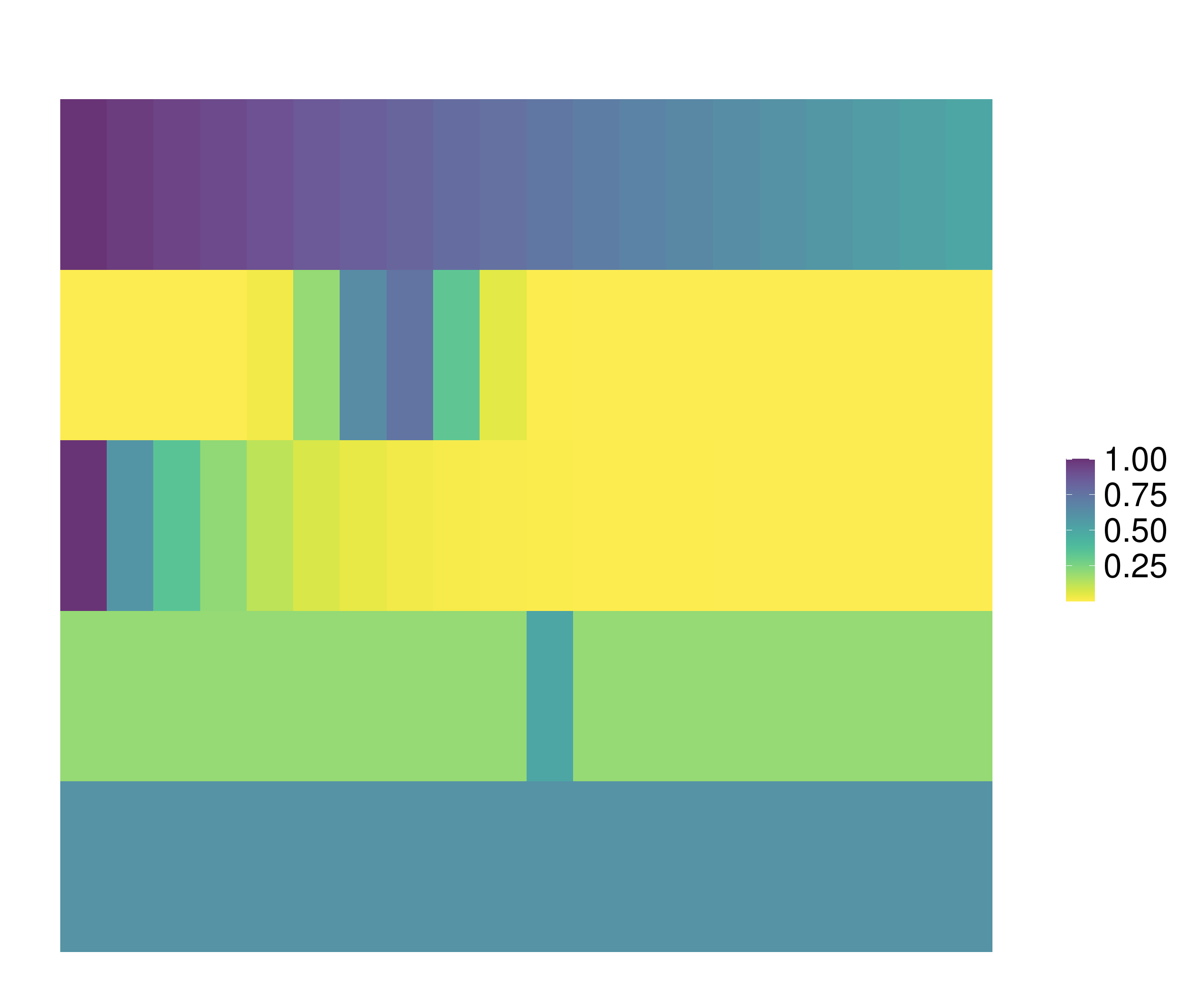}
          \end{minipage}
           \begin{minipage}{0.3\linewidth}
            \includegraphics[width = 0.95\linewidth]{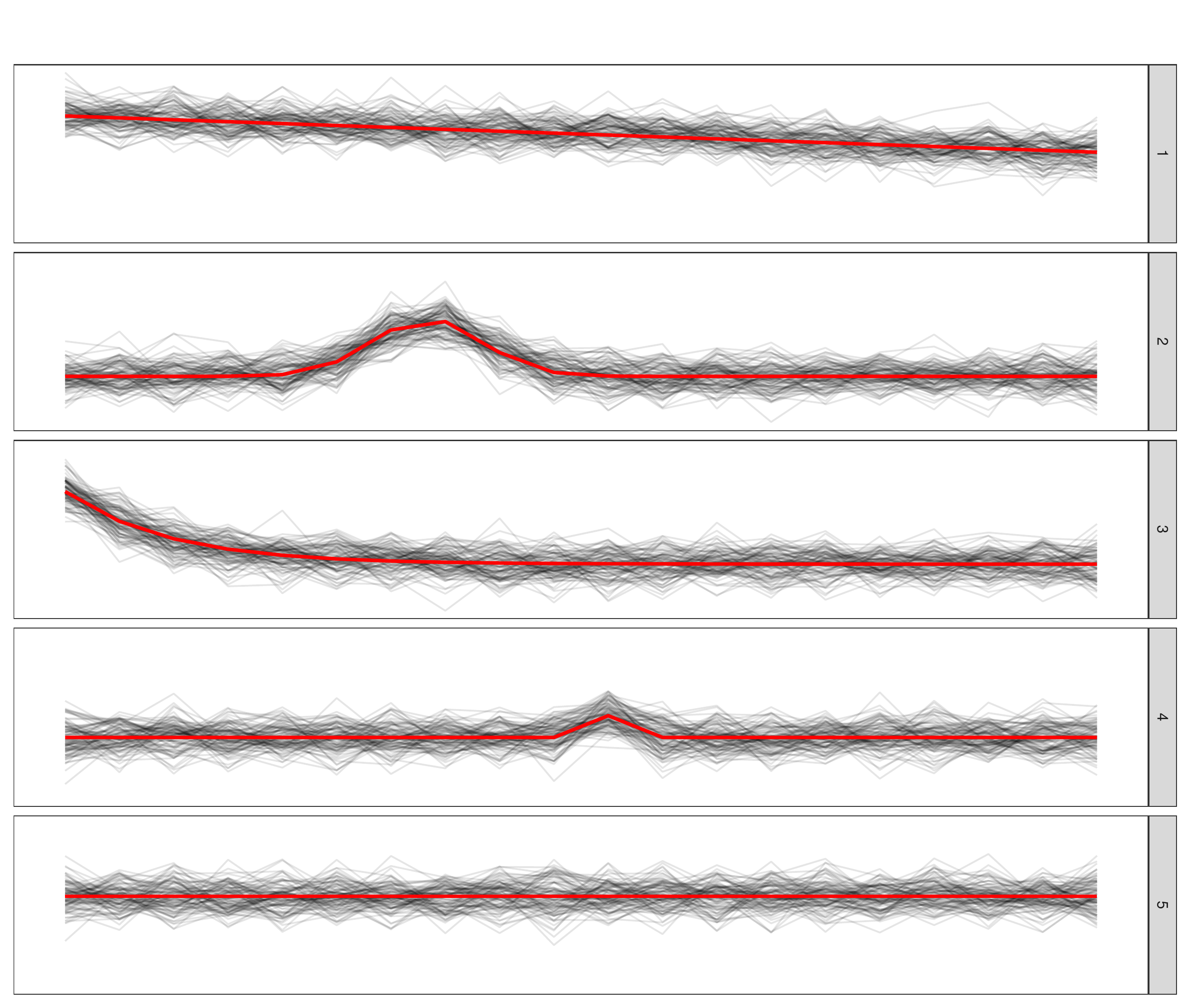}
      \end{minipage}%
\caption{Graphical representation of the three matrix prototypes used to define the groups in the various simulation settings. Cf. Fig. \ref{fig:centers} for the left and middle panels, associated to the settings $(P\times T) = (5\times 5)$ and $(P\times T) = (5 \times 20)$ respectively. The right panels display, for the $P=5$ variables of each cluster prototype, a subsample of curves generated via DCT (black dashed lines) with $\rho = 1$ and over-imposed the associated prototype of each row (red solid line).}
\label{fig:centers}
\end{figure*}

Three main clustering configurations are considered: a single-group setting, defined by the prototype $A$ illustrated in Figure \ref{fig:centers}; a balanced two-groups setting, with matrix-variate data equally sampled from prototypes $B$ and $C$ of Figure \ref{fig:centers}, and an imbalanced two-groups setting, with data again sampled from prototypes $B$ and $C$ in the uneven proportion $0.1$ and $0.9$, respectively. 
For each of these settings, varying sample size, data dimensions, cluster variability and distribution are evaluated, by letting $N \in \{1000, 3000\}$, $(P\times T) \in \{5\}\times \{5, 20\}$, and $\rho \in \{0.1,0.3,1\}.$ For each setting, 500 Monte Carlo samples have been generated. 

Modal clustering is performed via the mean-shift algorithm discussed in Section \ref{sec:mean-shift}, applied to three different formulations of kernel estimator. A fixed bandwidth estimator is evaluated, with Normal matrix-variate kernel and scalar bandwidth set as asymptotically optimal to estimate the first derivative of a Normal matrix-variate density. While this choice is unarguably sub-optimal, especially in the presence of multimodal structures, it has been proven successful in many applications of modal clustering in the standard multivariate setting \citep{menardi2016}. In fact, since this rule of thumb is known to oversmooth the true density, it seems in principle a sensible choice in the presence of high dimensional data, where oversmoothing may relieve the problem of spurious cluster in the low density regions. As a representative of balloon estimators, we consider the \eqref{eq:knn_b} with Uniform Kernel on the $PT-$ball of radius $\delta_k(\cdot),$ and $k \in (0.5\sqrt{N}, \sqrt{N}, 5\sqrt{N})$. 
Finally, we consider a sample point estimator \eqref{eq:knn_sp}, with Normal matrix-variate kernels, bandwidth $h \delta_{k}(X_n),$ $k \in (0.5\sqrt{N}, \sqrt{N}, 5\sqrt{N})$ and $h$ set as in the fixed bandwidth case. 
  
As a benchmark, we also perform clustering via $K$-means and model-based clustering based on mixtures of matrix-variate Normal distributions \citep{viroli2011model}. In the former case the number of clusters is determined by using the best \textit{Silhouette} score \citep{silhouette} in the range of values $\{2,\ldots, 9\}$, whereas in the latter case the BIC is computed in the range $\{1,\ldots, 9\}.$ 

The quality of the detected clustering is evaluated by comparing it with the true one via the Fowlkes–Mallows index \citep[see, e.g.][Ch. 27]{hennig_etal2015}, as it is sensitive to a different quality of partitions also when one of the two partitions is formed by one group only. 

All the analyses have been run in the \texttt{R} environment \citep{R}, with the modal clustering routines built as suitable modifications of functions available in the \texttt{ks} packages \citep{ks}, and the aid of the packages \texttt{stats} and \texttt{mclust} \citep{mclust} for running $K$-means and, respectively, model-based clustering.  

\subsection{Results} 

\begin{figure*}[t]
    \centering
    \begin{minipage}{\linewidth}
\hspace{.05\textwidth}
        \textsf{N = 1000, P = 5, T = 5}\par
    \hspace{.05\textwidth}
     \textsf{$\rho = 0.1$}
     \hspace{.23\textwidth}      \textsf{$\rho = 0.3$}
     \hspace{.22\textwidth}      \textsf{$\rho = 1$}
     \par
        \includegraphics[width =0.36 \linewidth, height=0.263\linewidth]{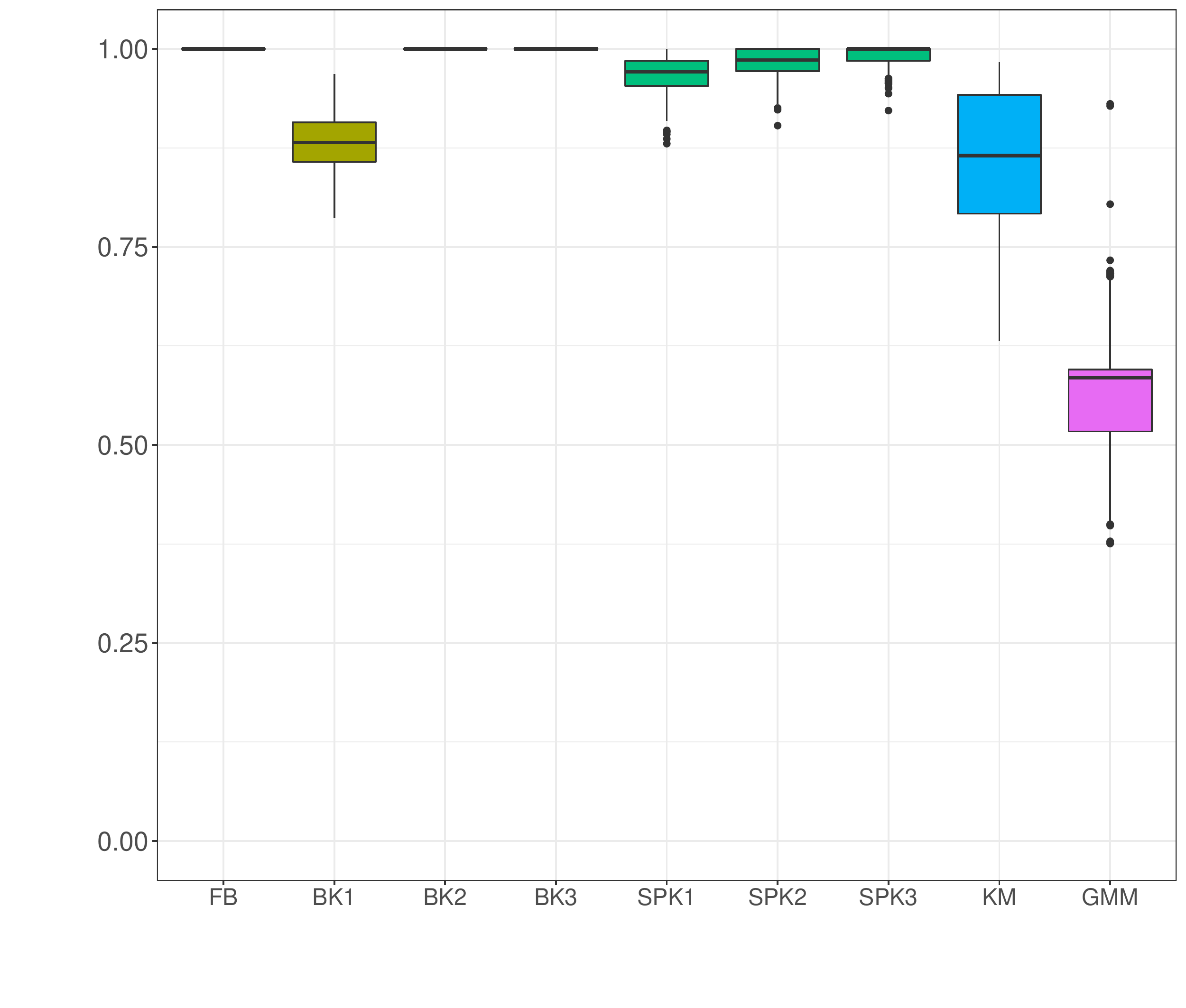}
            \includegraphics[width =0.31 \linewidth]{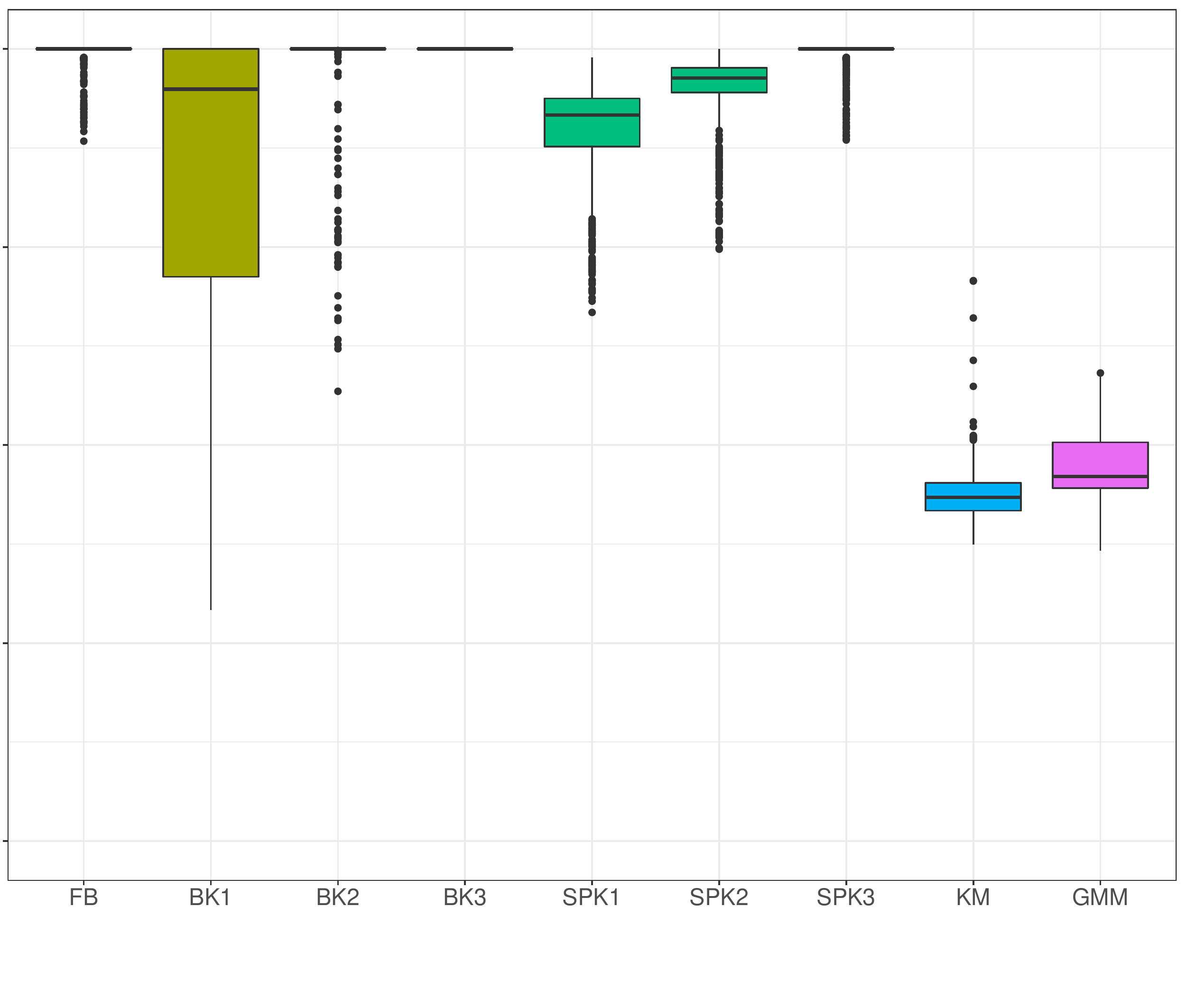}
                    \includegraphics[width =0.31 \linewidth]{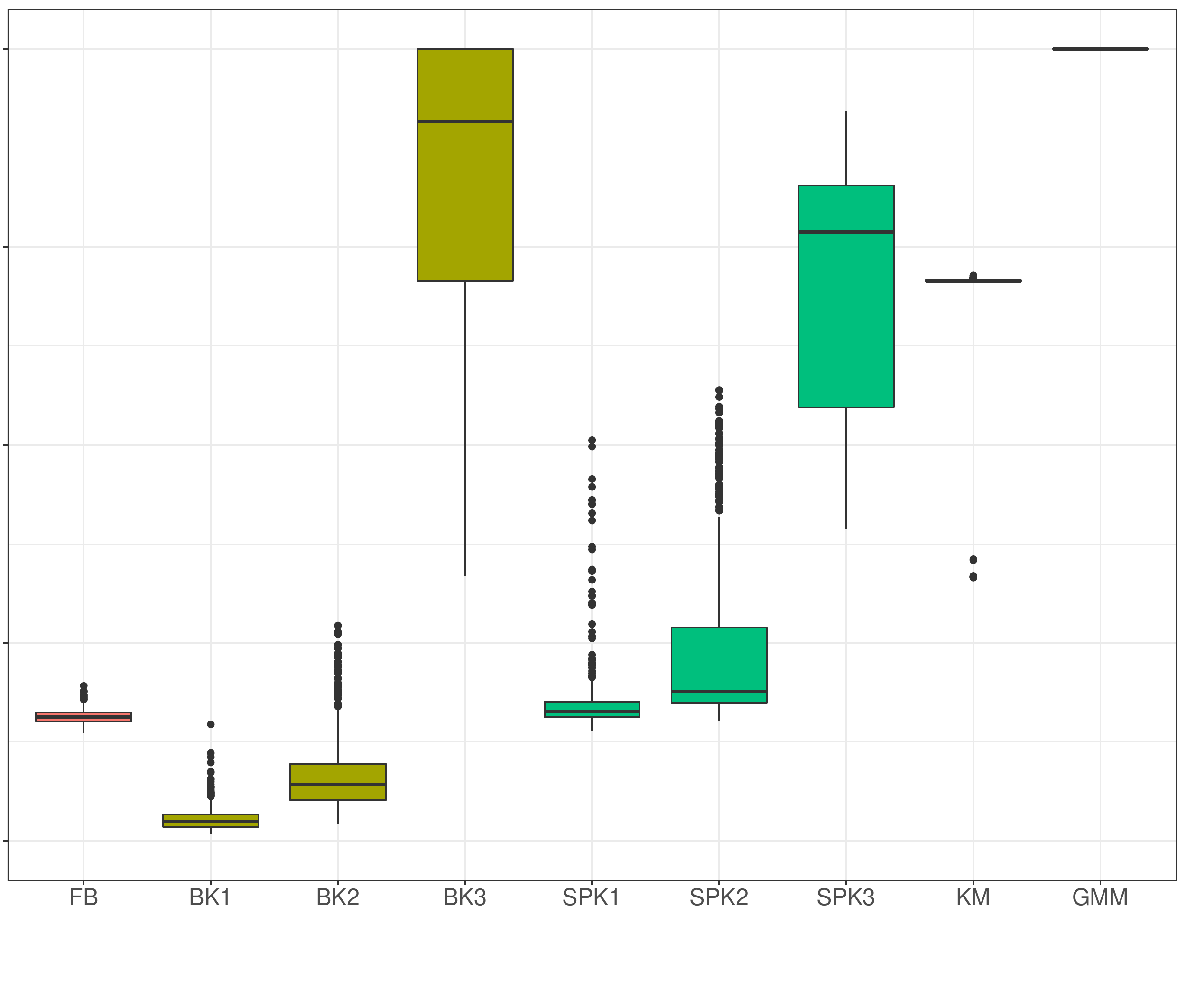}
    \end{minipage}
    \begin{minipage}{\linewidth}
\hspace{.05\textwidth}
        \textsf{N = 1000, P = 5, T = 20}\par
    \hspace{.05\textwidth}
     \textsf{$\rho = 0.1$}
     \hspace{.23\textwidth}      \textsf{$\rho = 0.3$}
     \hspace{.22\textwidth}      \textsf{$\rho = 1$}
     \par
         \includegraphics[width =0.36 \linewidth, height=0.263\linewidth]{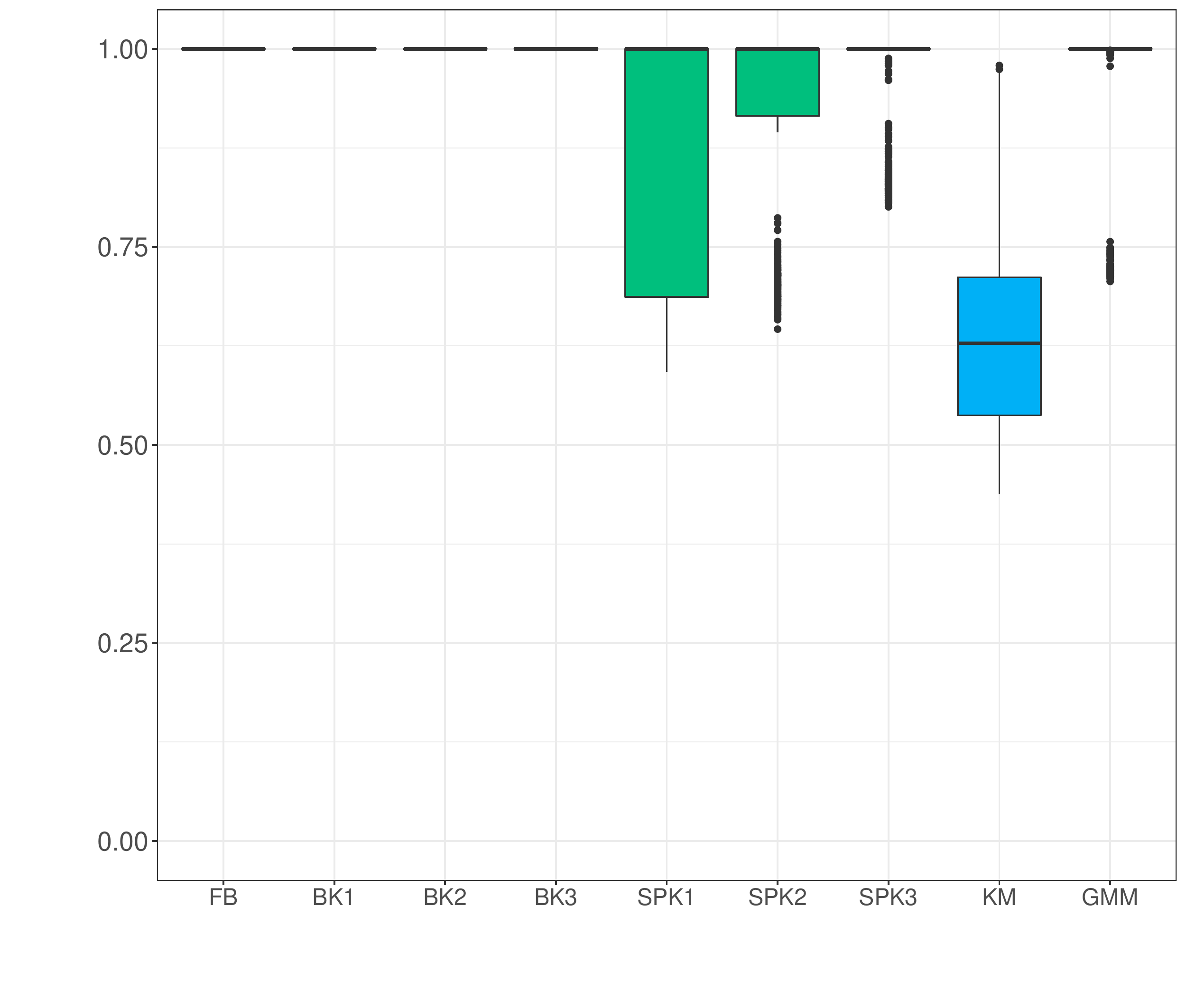}
            \includegraphics[width =0.31 \linewidth]{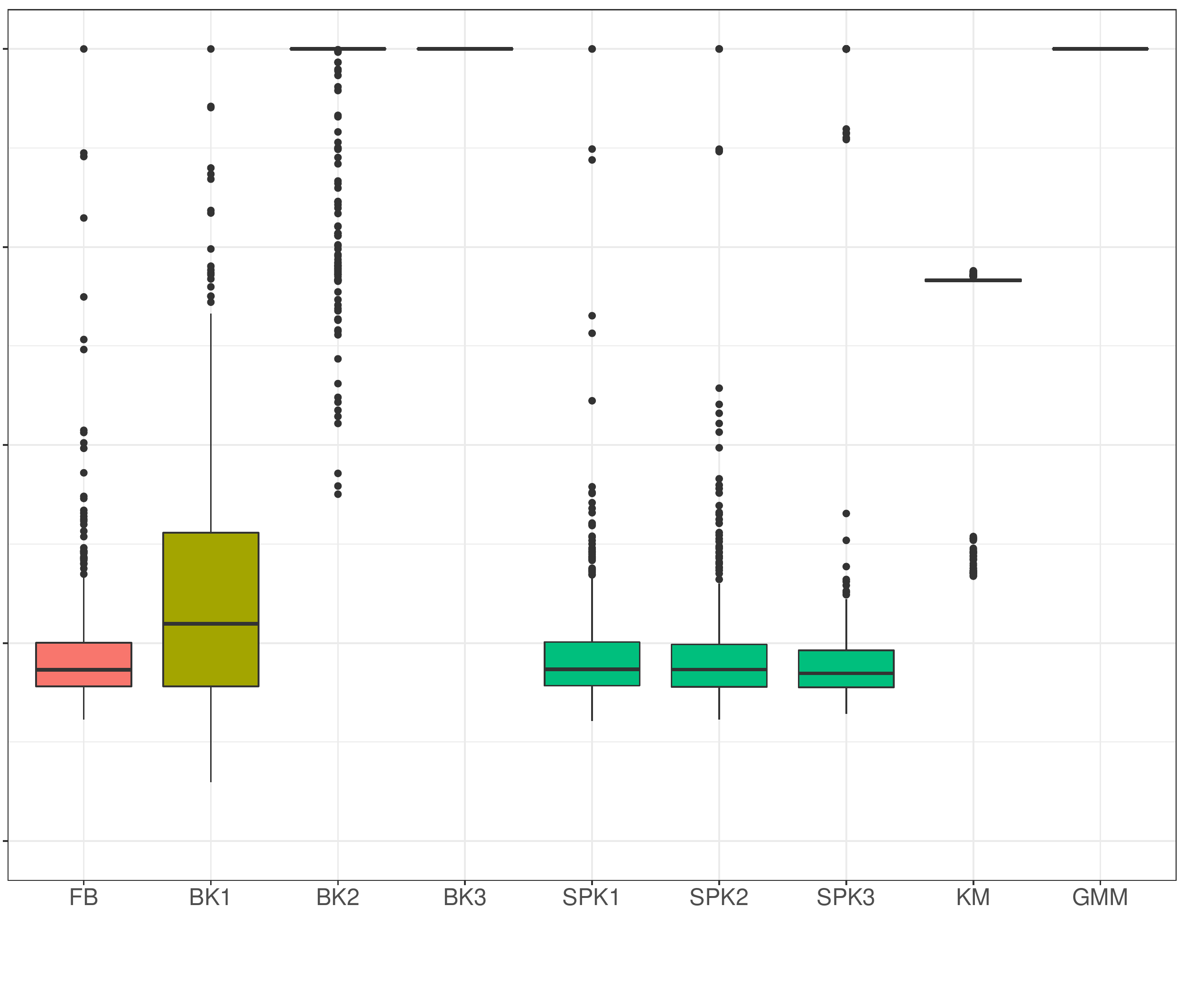}
                    \includegraphics[width =0.31 \linewidth]{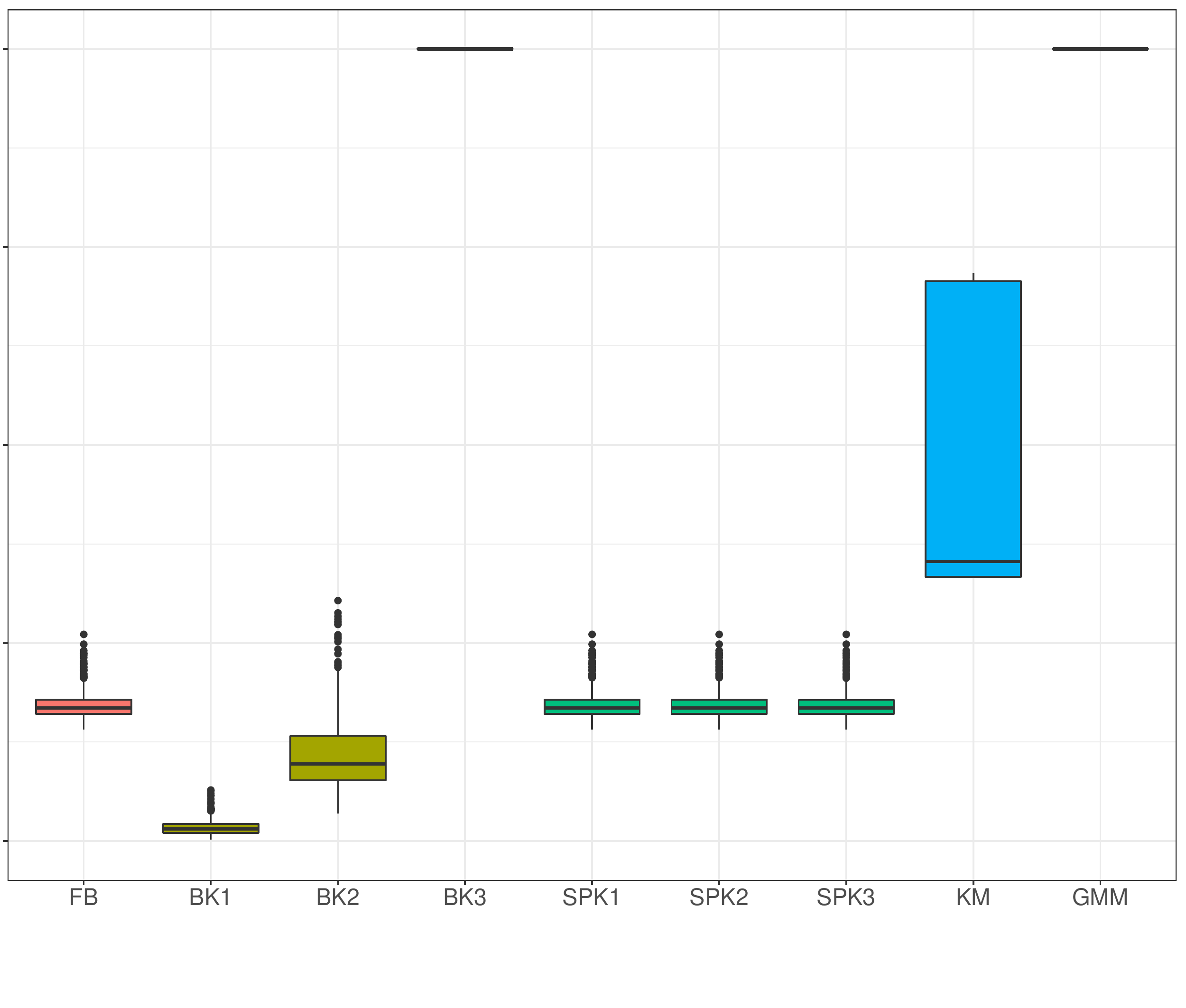}
    \end{minipage}
         \caption{Simulation results of the single group setting, with sample size, matrix dimension and proportion $\rho$ of DCT coefficients defined in the figure headings. Each panel displays the Monte Carlo distribution of the FM index when modal clustering is run with a fixed bandwidth kernel estimator (FB), a balloon and a sample point $k$-NN estimator, both with increasing values of $k$ (BK1, BK2, BK3, SPK1, SPK2, SPK3), and when $K$-means and model-based clustering with gaussian mixture models are run (KM and, respectively, GMM).}
\label{fig:single}
\end{figure*}

\begin{figure*}[t]
    \centering
    \begin{minipage}{\linewidth}
    \hspace{.05\textwidth}
        \textsf{N = 1000 5x5}\par
        \hspace{.05\textwidth}
     \textsf{$\rho = 0.1$}
     \hspace{.23\textwidth}      \textsf{$\rho = 0.3$}
     \hspace{.22\textwidth}      \textsf{$\rho = 1$}
     \par
         \includegraphics[width =0.36 \linewidth,height=0.263\linewidth]{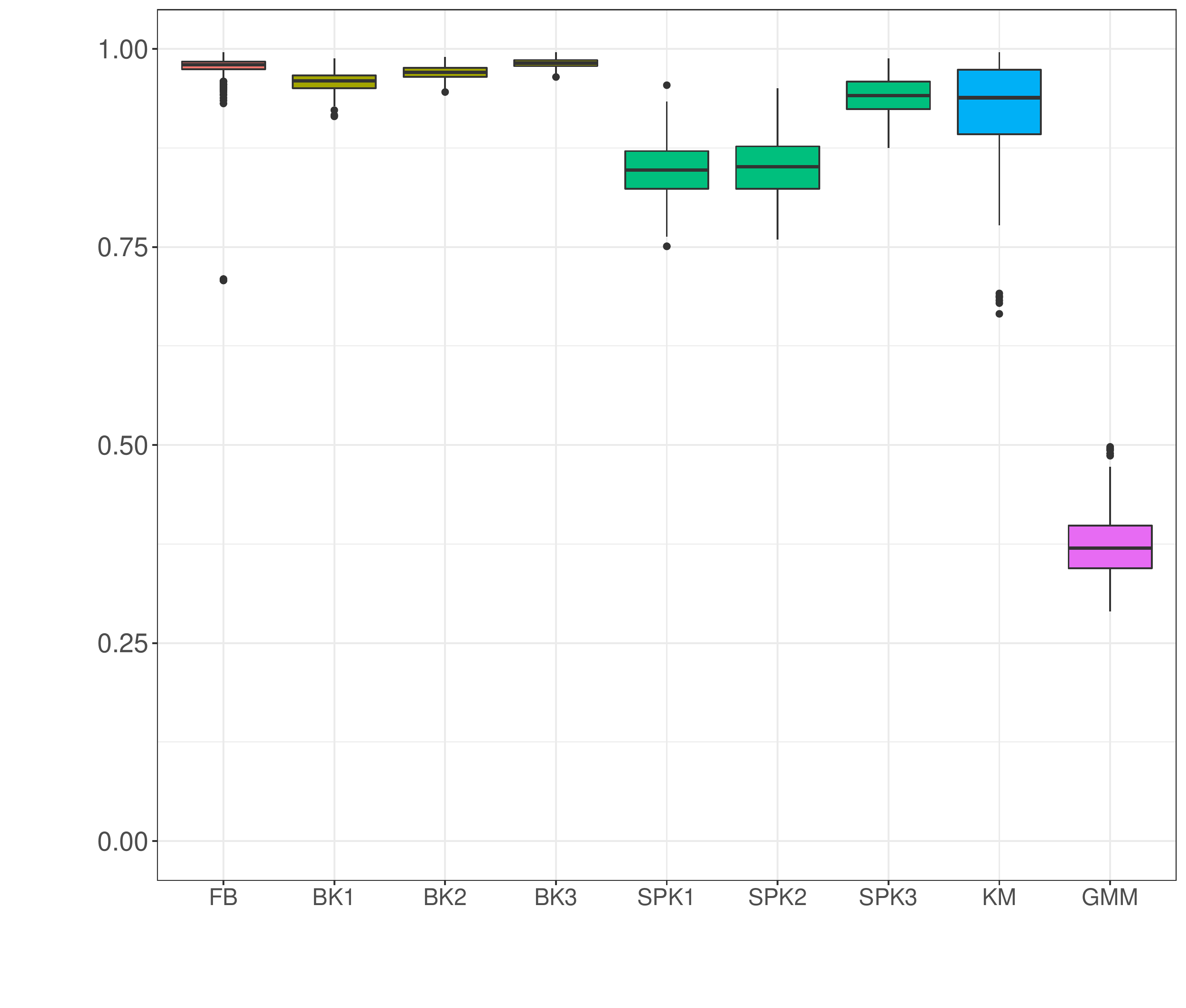}
            \includegraphics[width =0.31 \linewidth]{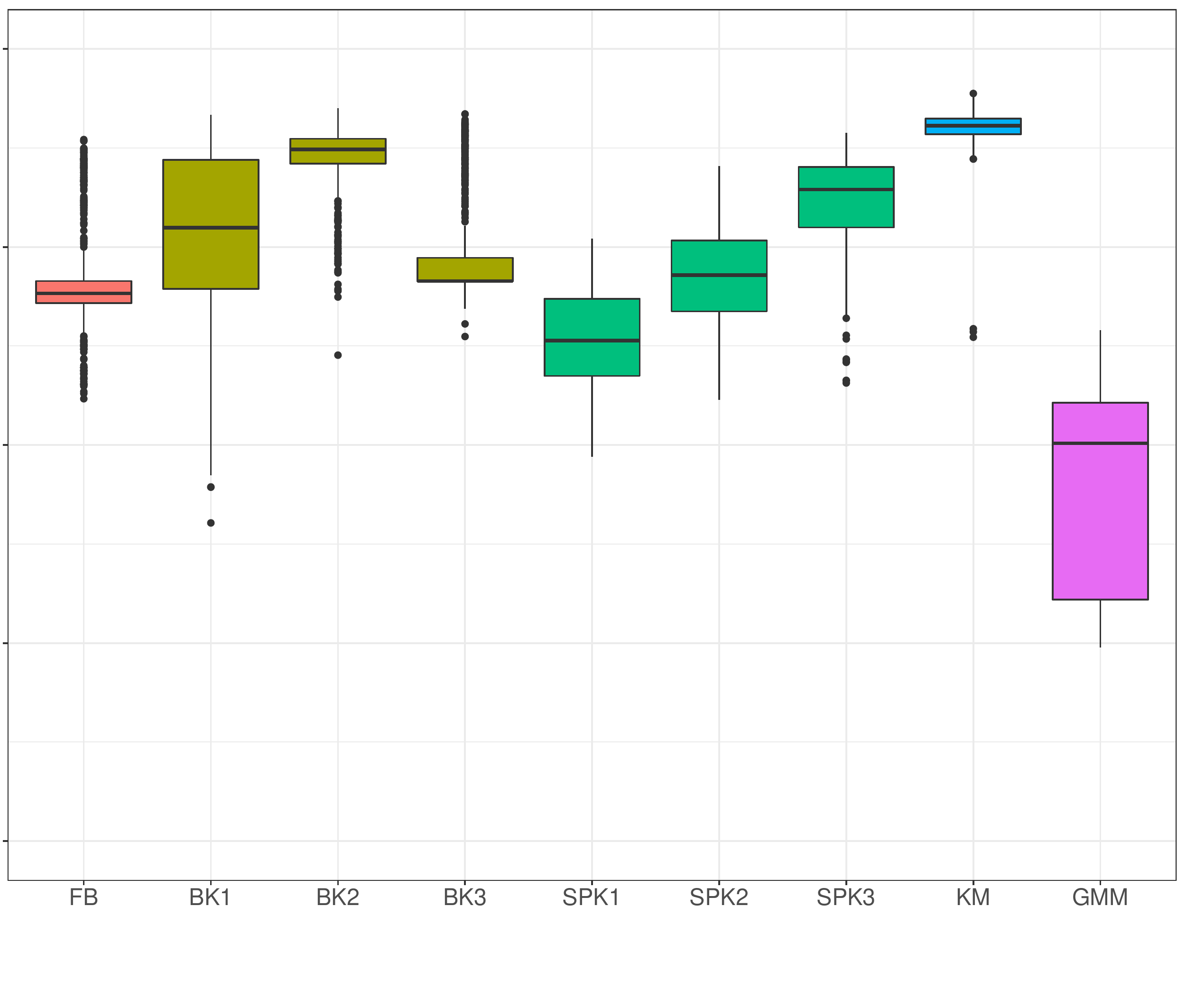}
                    \includegraphics[width =0.31 \linewidth]{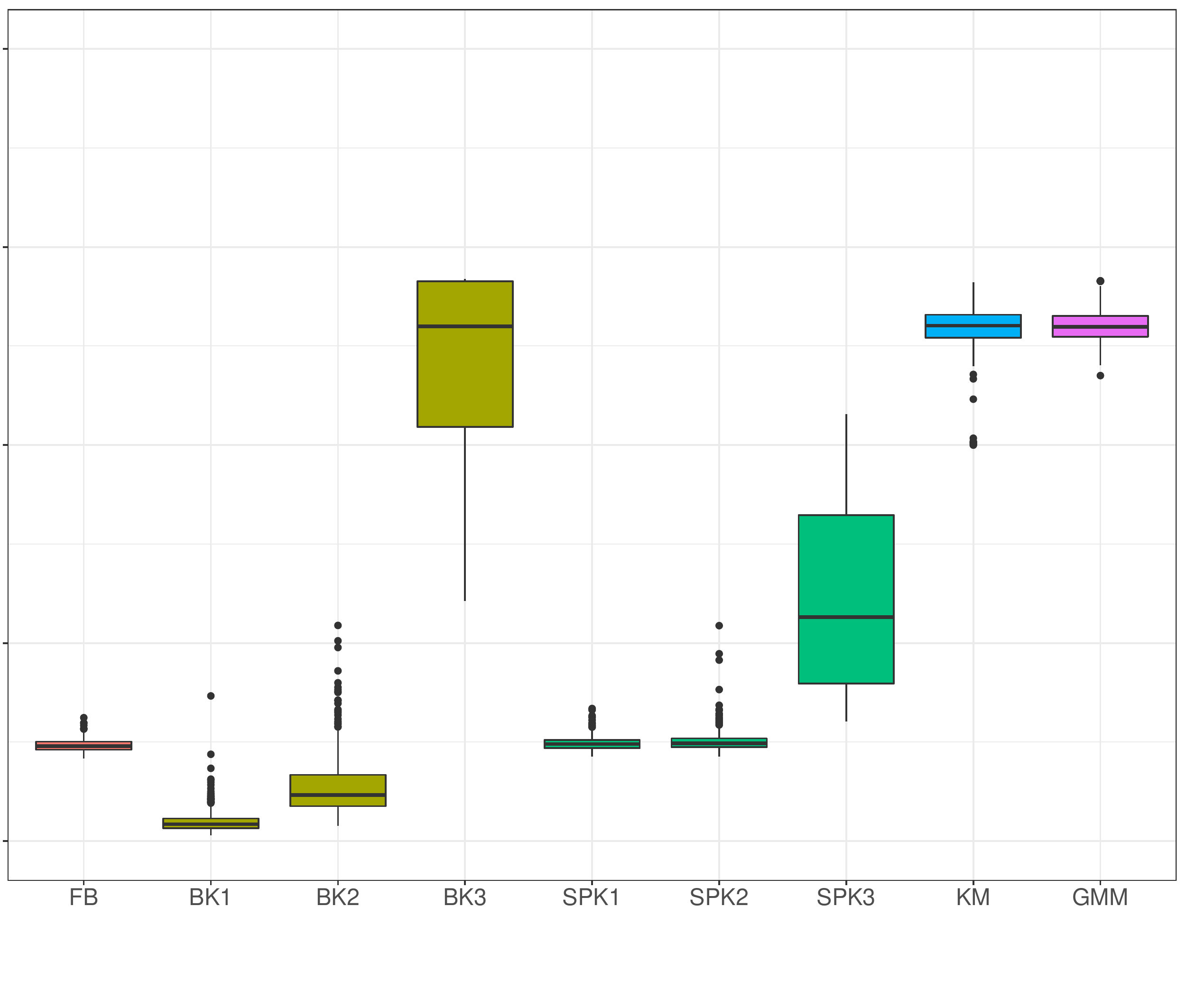}
    \end{minipage}
    \begin{minipage}{\linewidth}
    \hspace{.05\textwidth}
        \textsf{N = 1000 5x20}\par
        \hspace{.05\textwidth}
     \textsf{$\rho = 0.1$}
     \hspace{.23\textwidth}      \textsf{$\rho = 0.3$}
     \hspace{.22\textwidth}      \textsf{$\rho = 1$}
     \par
         \includegraphics[width =0.36 \linewidth, height=0.263\linewidth]{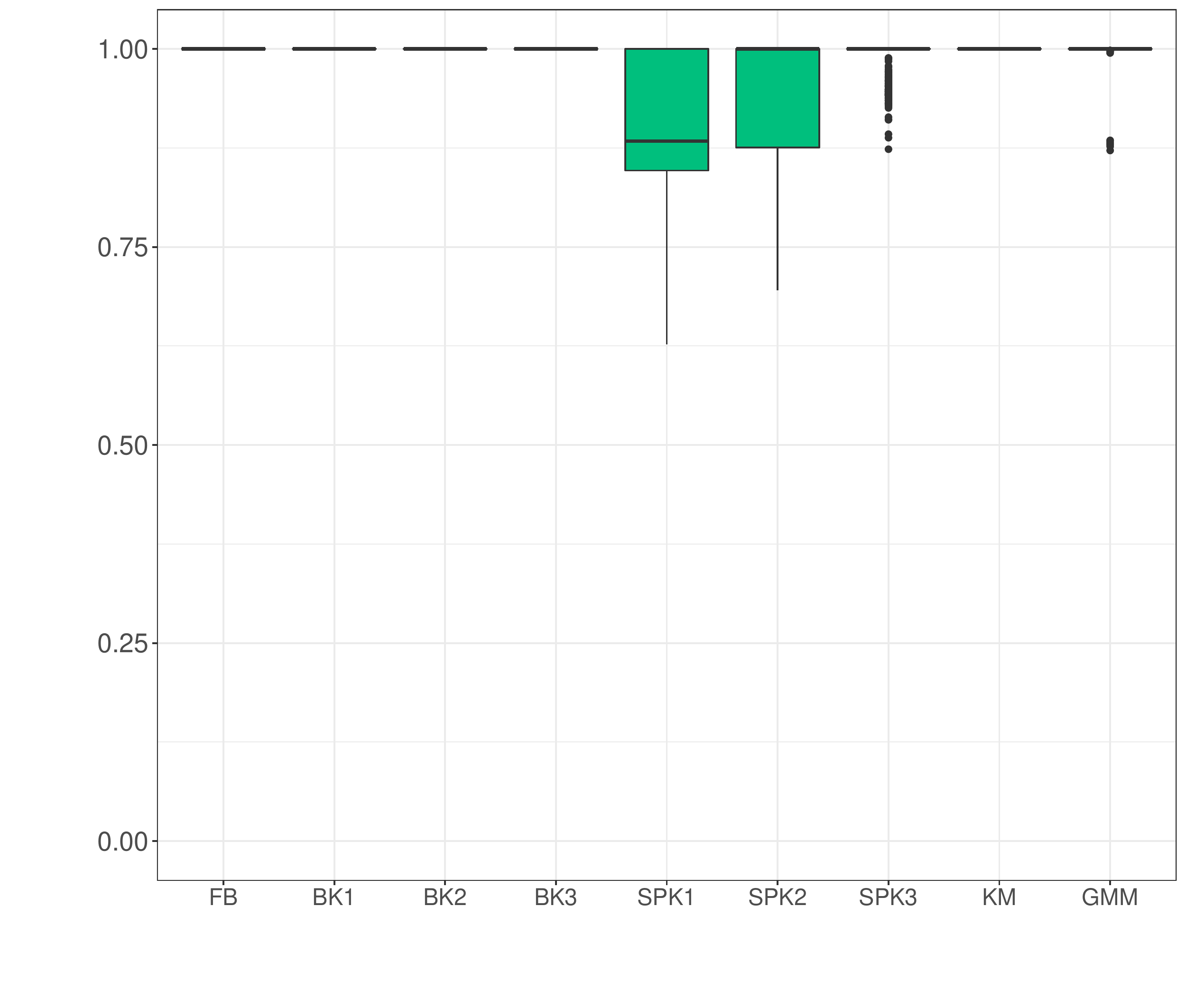}
            \includegraphics[width =0.31 \linewidth]{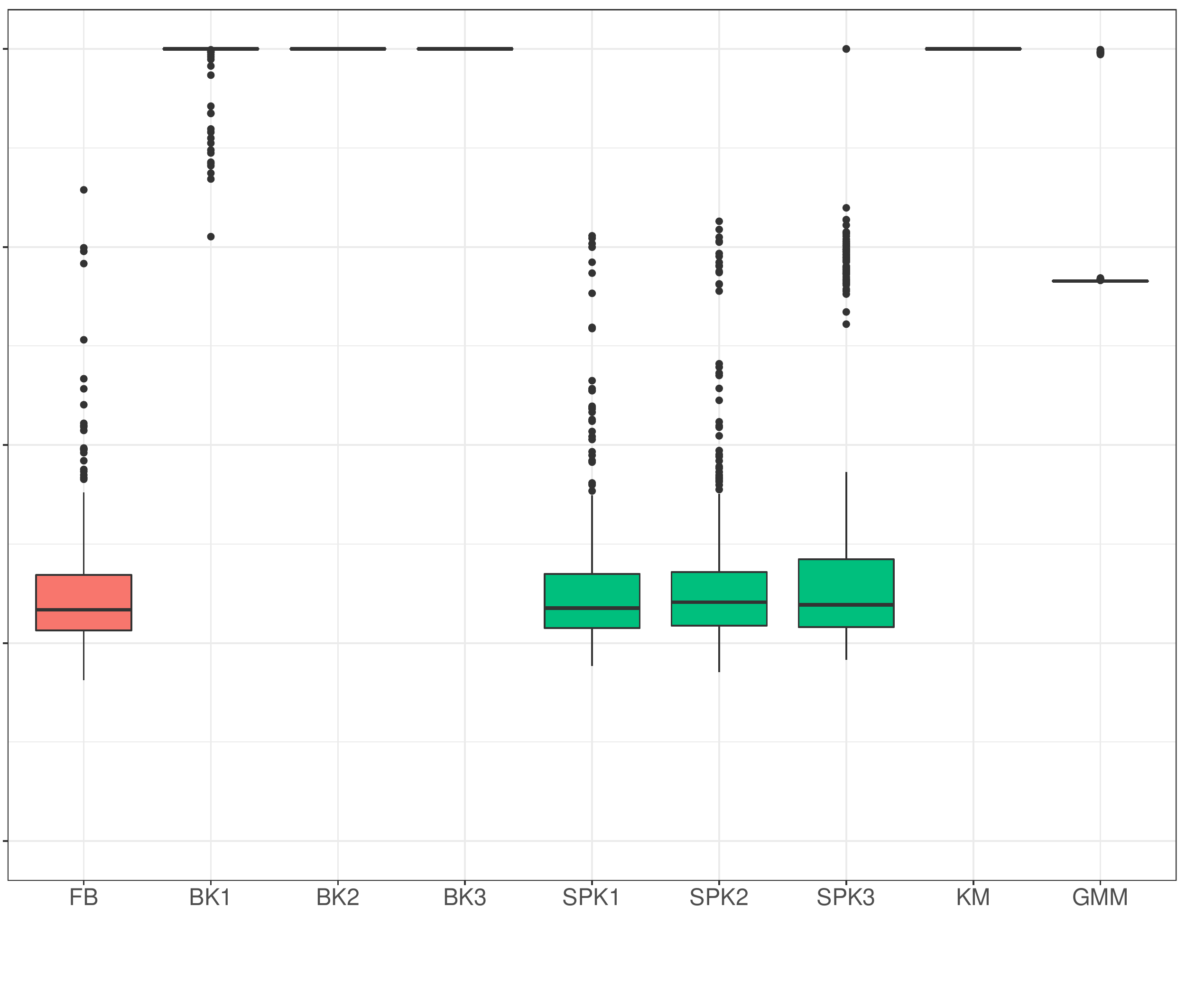}
                    \includegraphics[width =0.31 \linewidth]{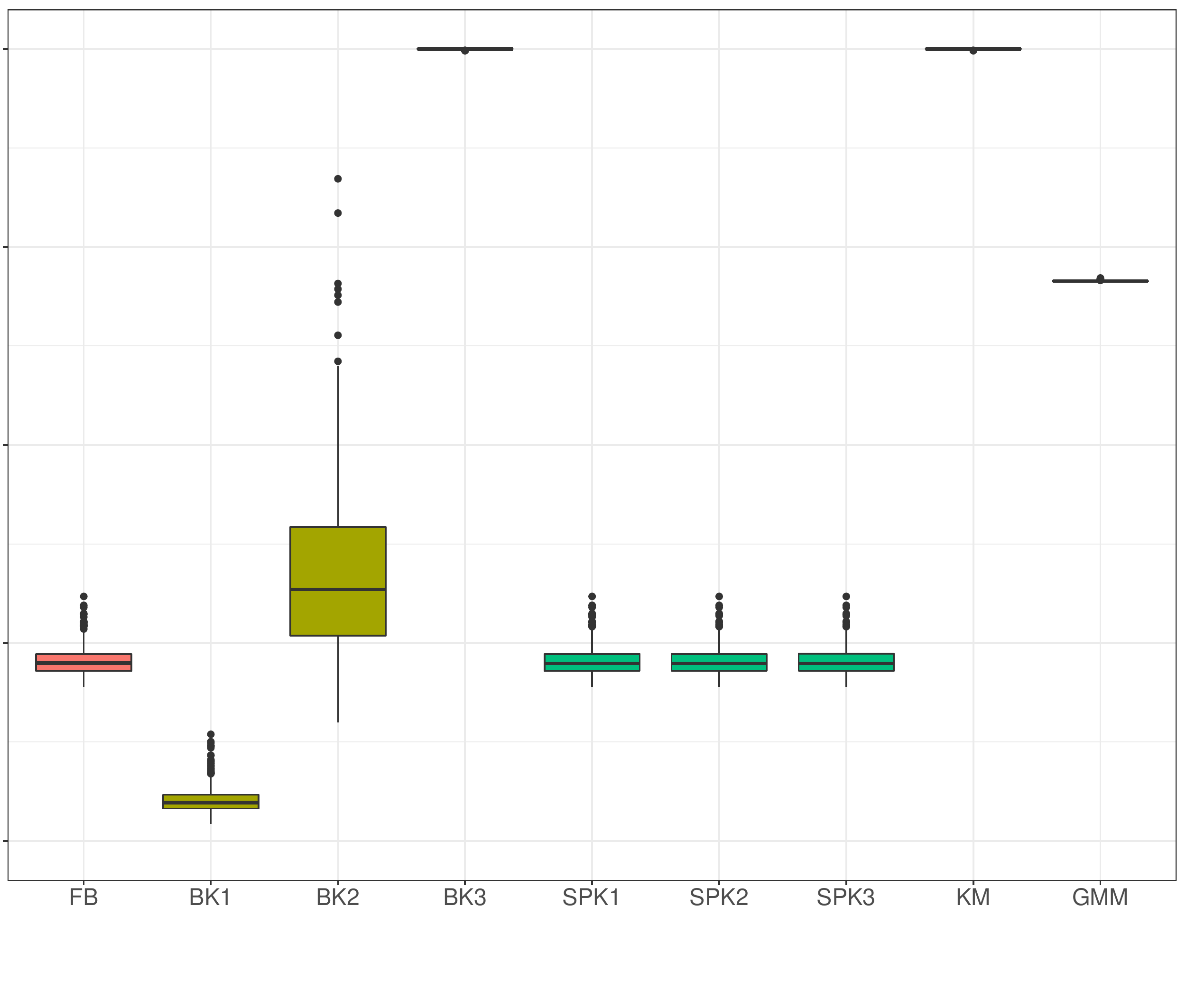}
    \end{minipage}
    \caption{Simulation results of the balanced groups settings. Cf Fig. \ref{fig:single}.}
\label{fig:bal}
\end{figure*}

\begin{figure*}[t]
    \centering
    \begin{minipage}{\linewidth}
    \hspace{.05\textwidth}
        \textsf{N = 1000 5x5}\par
        \hspace{.05\textwidth}
     \textsf{$\rho = 0.1$}
     \hspace{.23\textwidth}      \textsf{$\rho = 0.3$}
     \hspace{.22\textwidth}      \textsf{$\rho = 1$}
     \par
                  \includegraphics[width =0.36 \linewidth, height=0.263\linewidth]{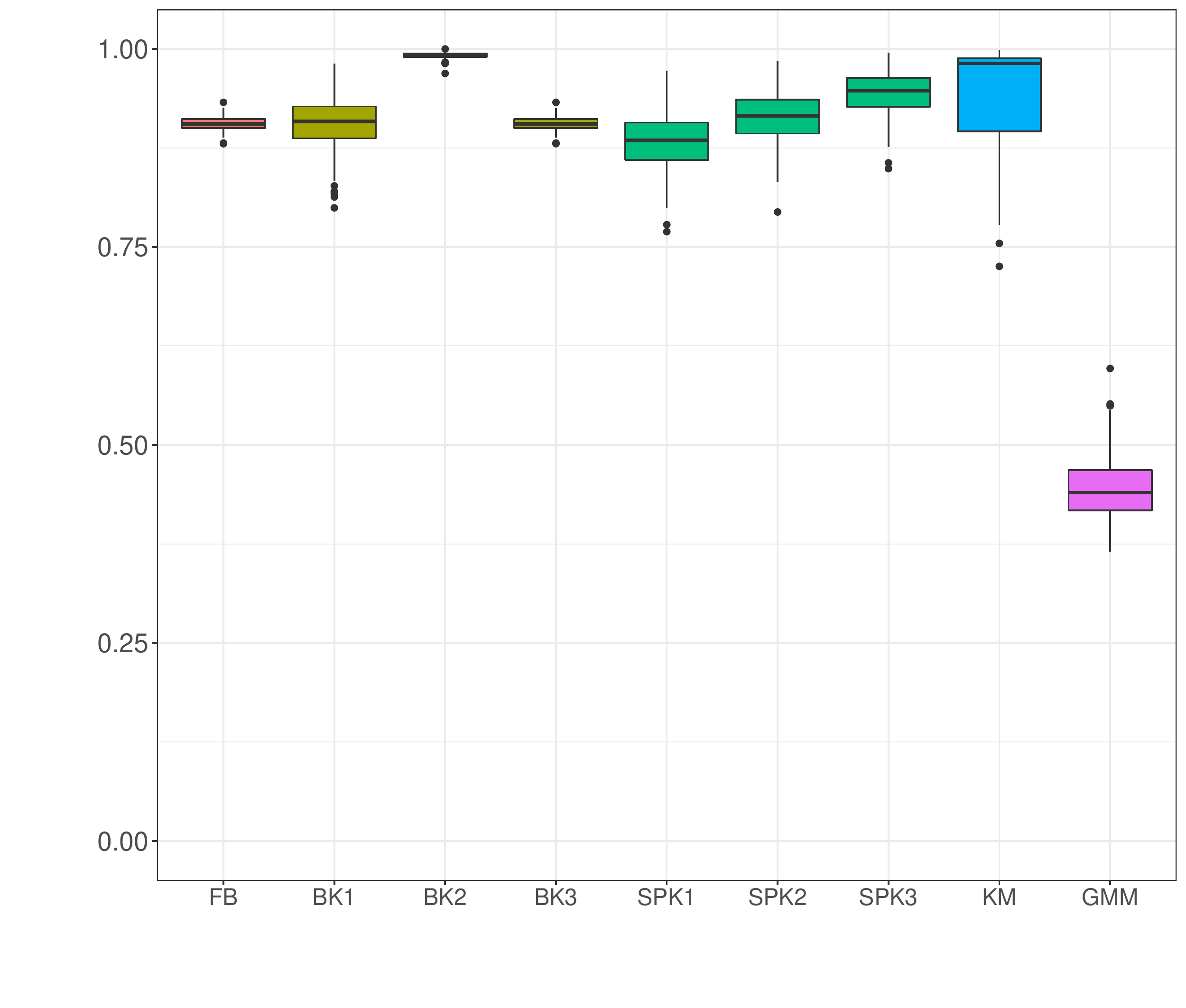}
            \includegraphics[width =0.31 \linewidth]{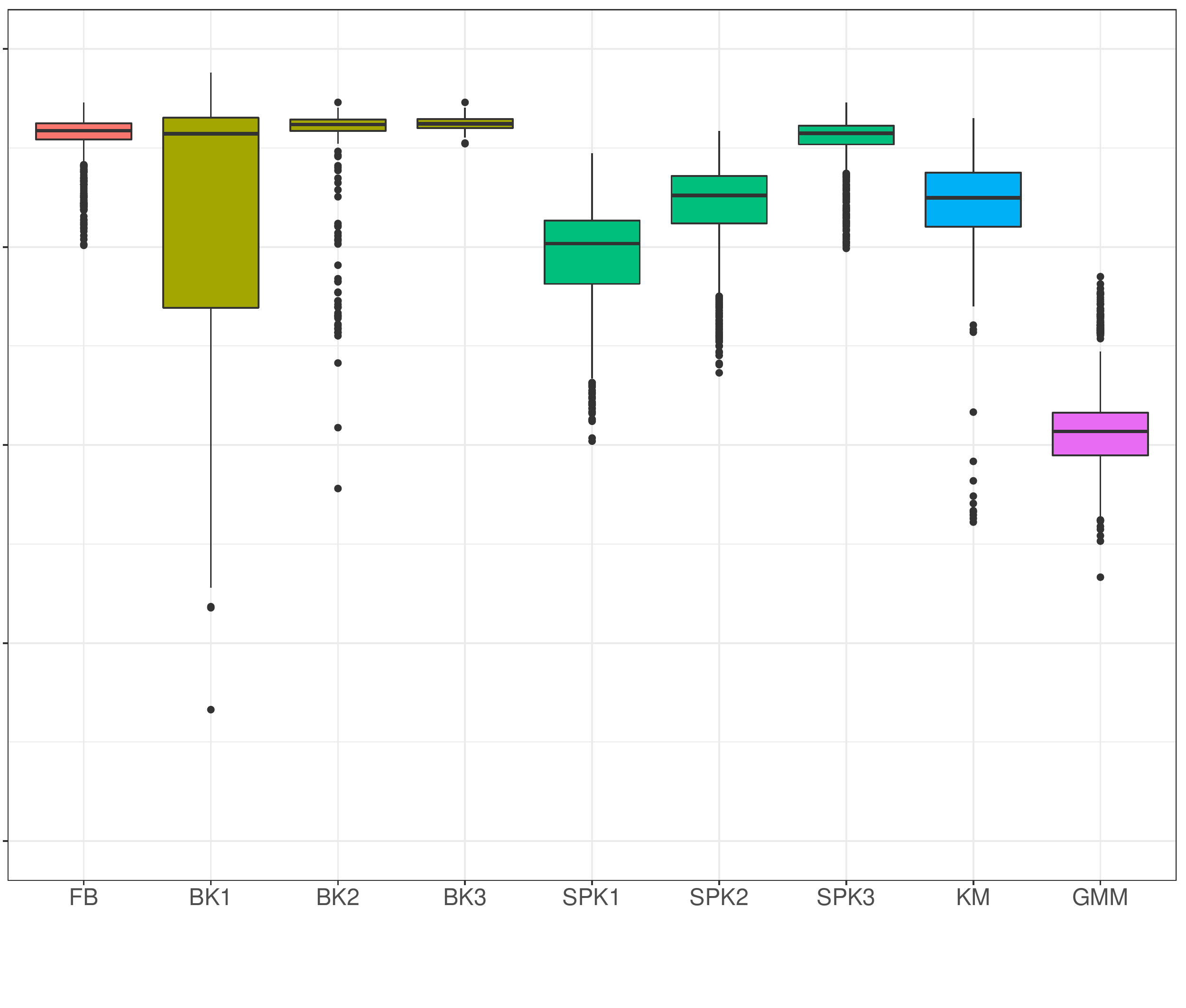}
                    \includegraphics[width =0.31 \linewidth]{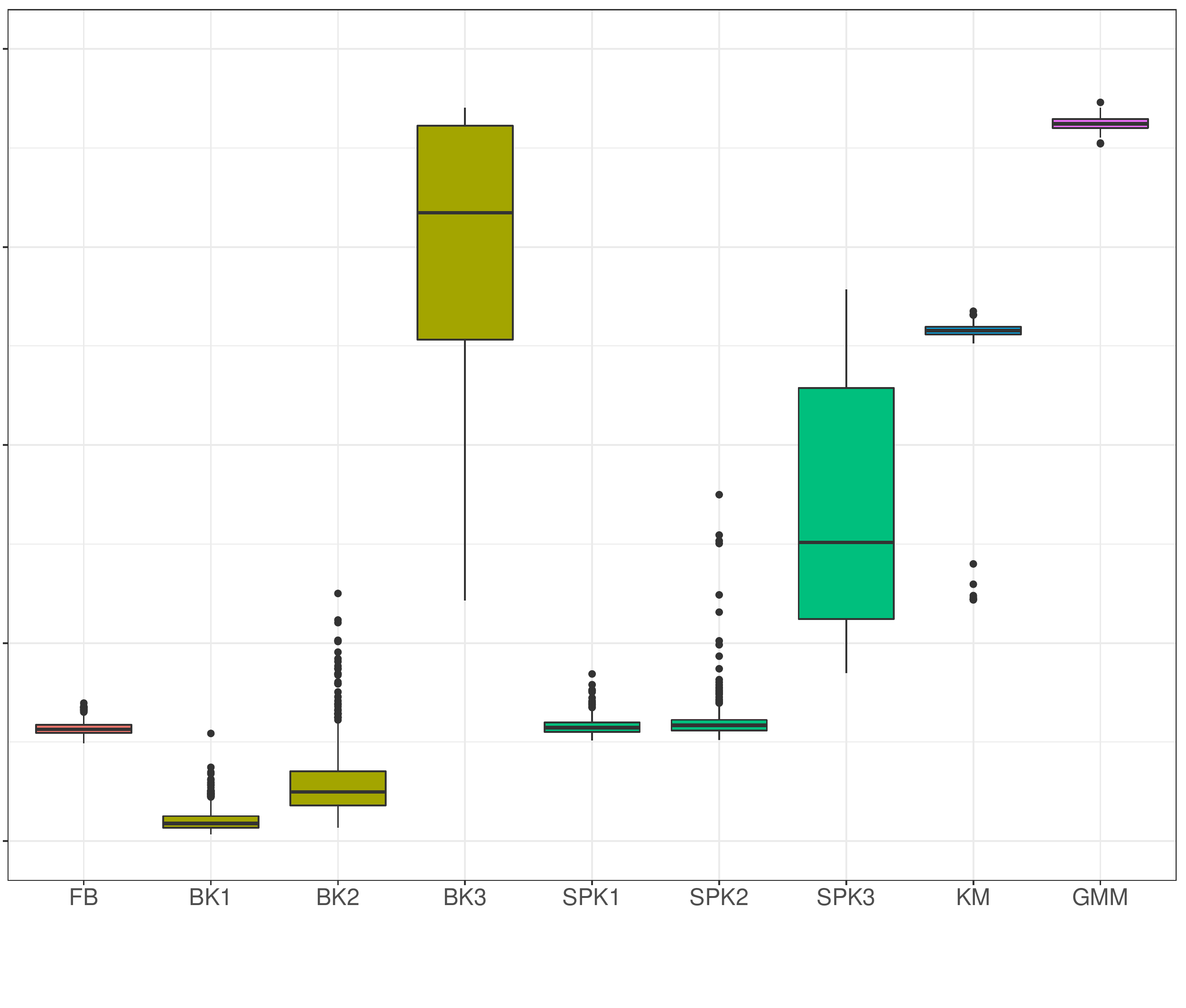}
    \end{minipage}
    \begin{minipage}{\linewidth}
    \hspace{.05\textwidth}
        \textsf{N = 1000 5x20}\par
        \hspace{.05\textwidth}
     \textsf{$\rho = 0.1$}
     \hspace{.23\textwidth}      \textsf{$\rho = 0.3$}
     \hspace{.22\textwidth}      \textsf{$\rho = 1$}
     \par
               \includegraphics[width =0.36 \linewidth, height=0.263\linewidth]{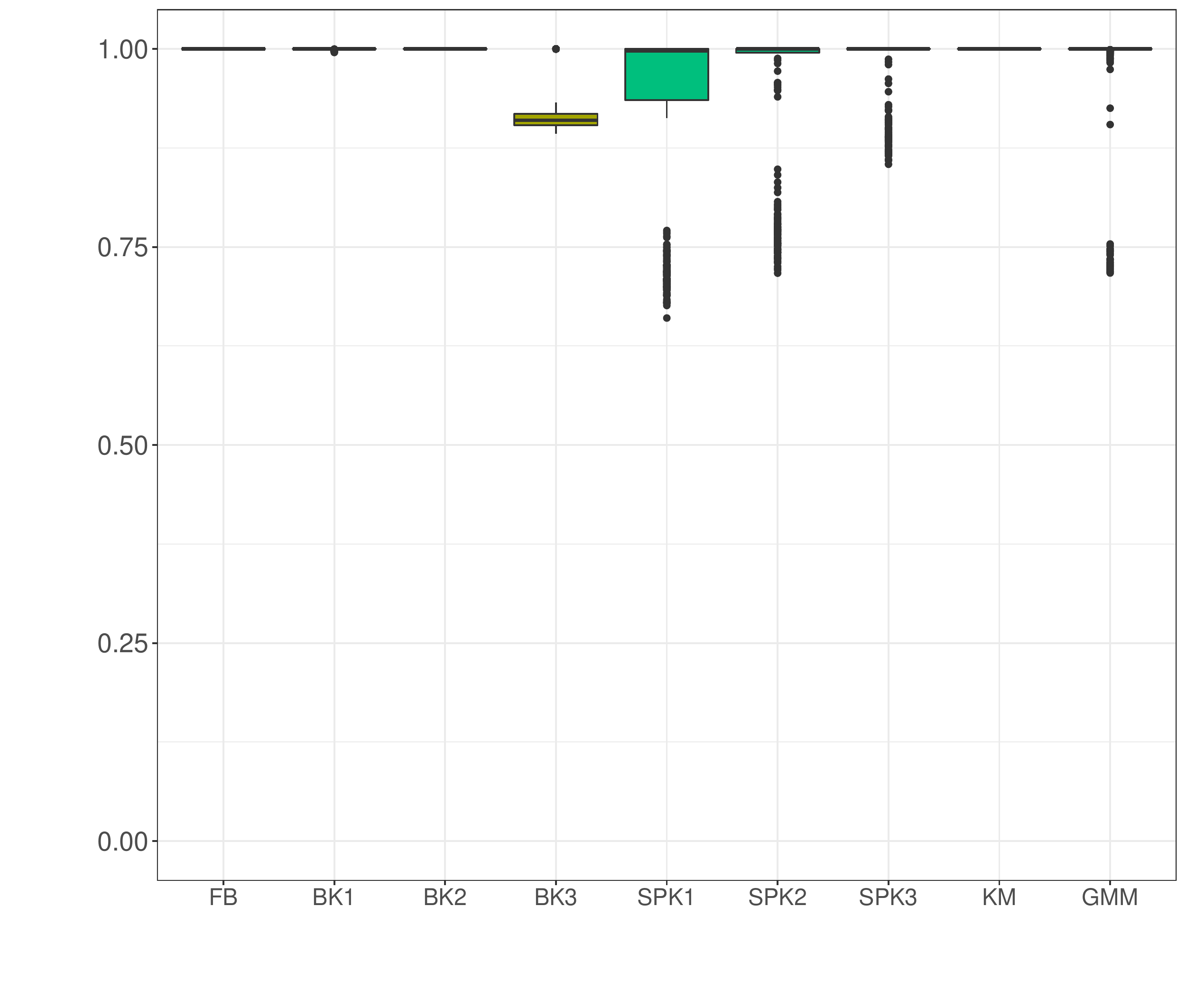}
            \includegraphics[width =0.31 \linewidth]{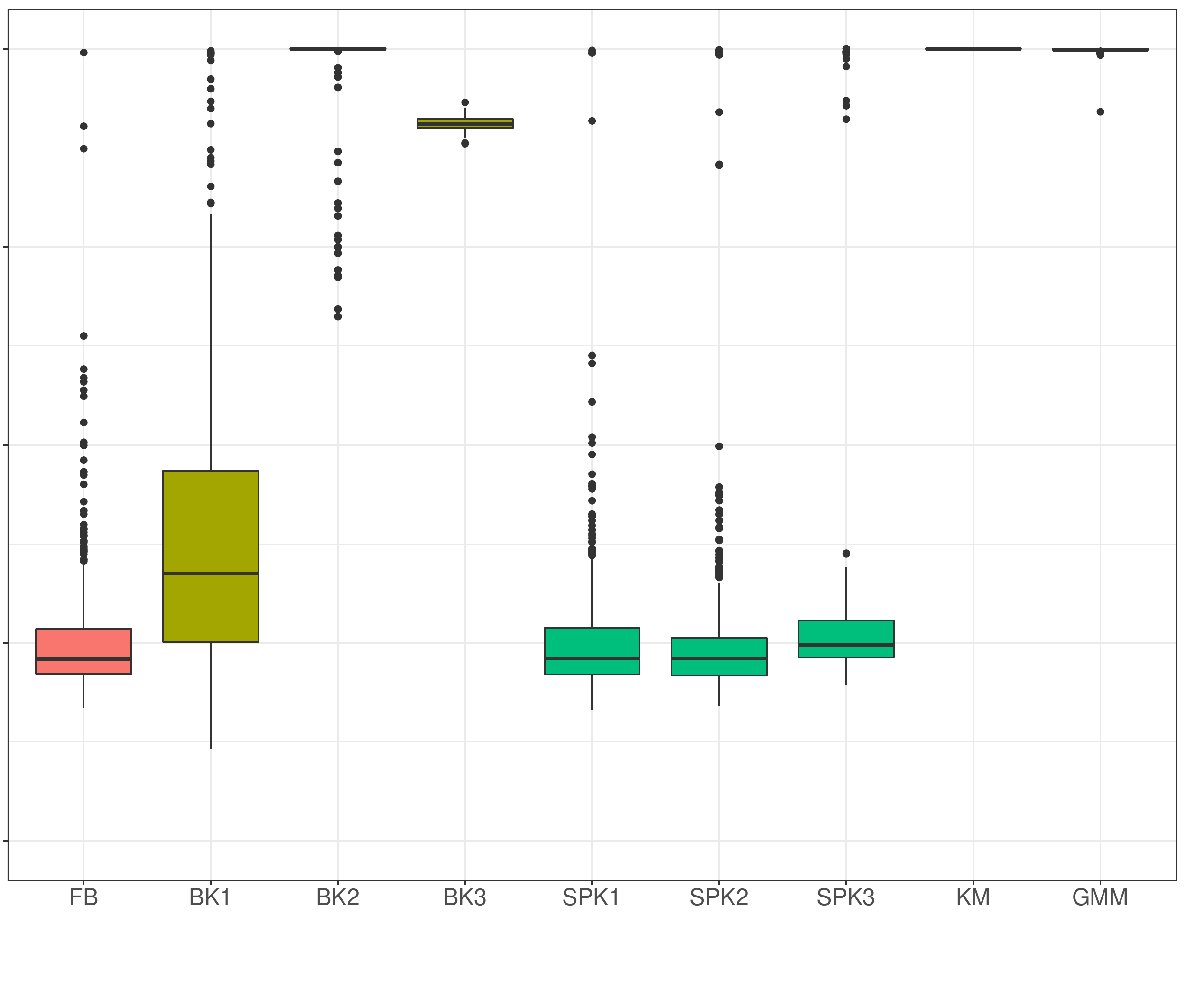}
                    \includegraphics[width =0.31 \linewidth]{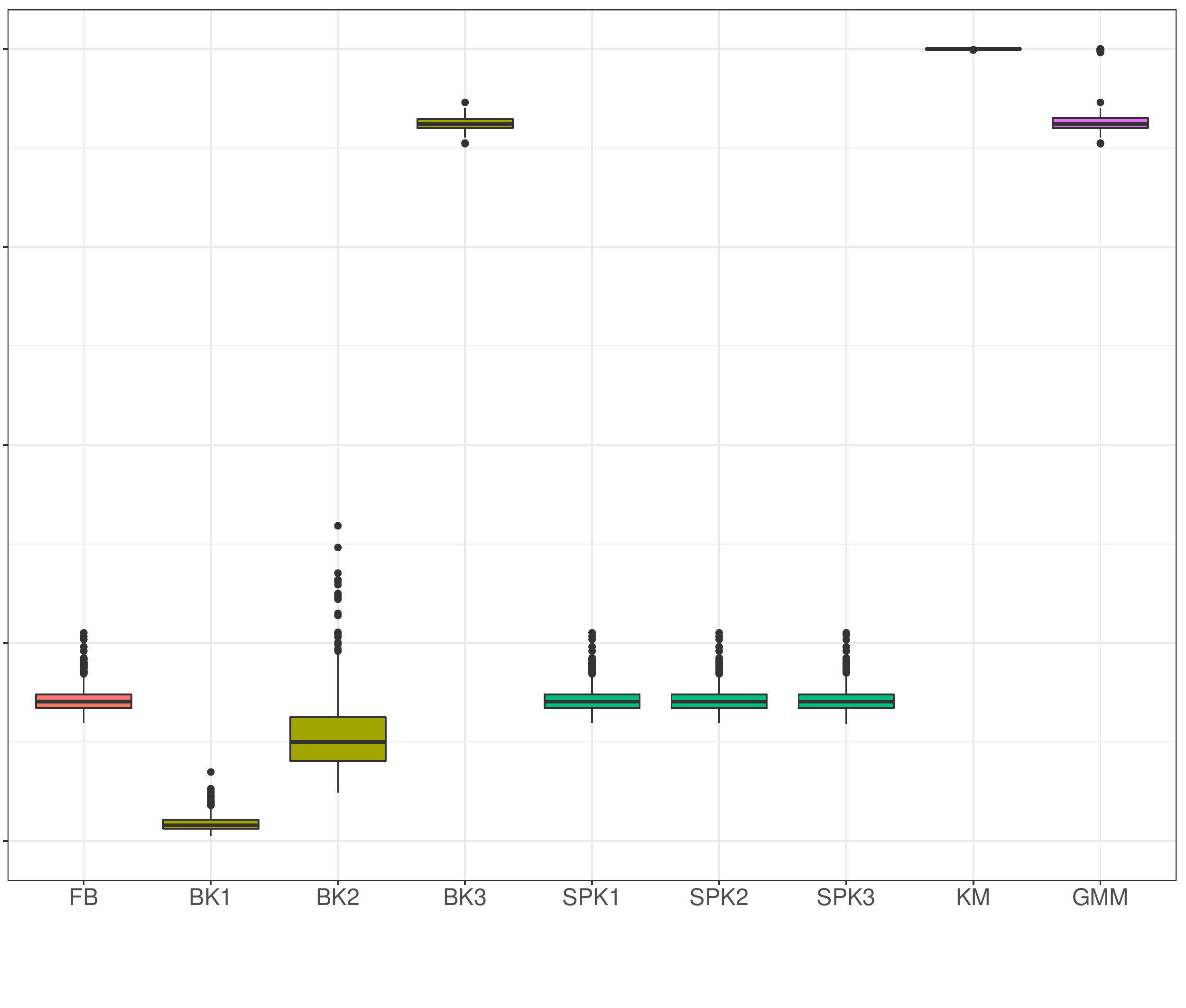}
    \end{minipage}
    \caption{Simulation results of the unbalanced groups settings. Cf Fig. \ref{fig:single}.}
\label{fig:unbal}
\end{figure*}

Results referred to simulations of samples of size $N=1000$ are displayed in Figures \ref{fig:single}, \ref{fig:bal}, \ref{fig:unbal}. Modal clustering performs successfully in all the considered settings, yet with some not negligible differences. 
The balloon $k$-NN kernel estimator is the one which mostly offers guarantees of revealing the true modal structure, as in all the considered settings there exists at least one value of $k$, among the examined ones, leading to a very accurate cluster detection. 
Cluster quality is not much sensitive to the selected number of nearest neighbors, at least for low to moderate amount of variability. For large $\rho,$ conversely, better results are achieved by a larger amount of smoothing, i.e. when $k$ takes its largest value among the three examined ones. It is worth to note that such largest value of $k$ produces an accurate classification of the observations in all the considered settings. 

Estimating the data density with the use of a scalar bandwidth, as well as via the sample point estimator, results in a faithful cluster recovery in the presence of a small amount of variability. Increasing $\rho,$ conversely, produces a progressive worsening of the results. A deeper insight of the results suggests that such lower accuracy is due to the arising of spurious clusters. 

It is perhaps unexpected that increasing the matrix dimension not always reduces the quality of detected clusters. In fact, the larger sparsity of the data in higher dimensions increases cluster separation, \emph{ceteris paribus}, hence the performance of modal clustering tends to improve. For those situations where this is not true, usually associated to the use of a sample point estimator, the arising of spurious clusters is the main responsible for the worsened behavior, hence we guess that a larger amount of smoothing would relieve such behaviour. Note, in fact, that in the considered estimators based on the $k$-nearest neighbors, the examined rules of thumb to select $k$ vary with the sample size and not with the data dimensionality.  
While further research should usefully shed light on specific criteria for bandwidth selection, we believe that relevant results have emerged from the analysis, confirming the opportunity of a satisfactory use of nonparametric density-based methods even in high dimensional spaces. 

Despite its simplicity and its known limitations, $K$-means clustering produces overall notable results. In the two-groups settings the quality of the detected partitions ranges from fair to very good, and the clustering structure is roughly caught even when the amount of variability is high. The algorithm finds it harder in the unbalanced settings, yet it can anyway identify the gross clustering structure. 
Not surprisingly, the worst results refer to the single cluster settings. In fact, the number of clusters set in $K$-means is selected by maximizing the Silhouette index, which cannot, by construction, be evaluated when $K=1.$ Hence, all the $K$-means results refer to partitions formed by at least two clusters. The low values of the Fowlkes-Mallows index, however, suggest that whatever number of clusters is selected by the Silhouette score, in the single-group settings observations are allocated uniformly to the clusters, instead of favouring a single group. This result is, in fact, consistent with the usual behavior of $K$-means clustering which tends to split data into balanced groups. This reason, along with the lack of a formal criterion to determine the number of clusters, overall discourage from the use of $K$-means, similarly to the standard multivariate framework. 

Model-based clustering is known to represent a generalization of $K$-means, where clusters are modeled to possibly vary in variance and proportion. Additionally, unlike $K$-means, the use of the BIC allows for selecting single-cluster models. Despite these advantages, model-based clustering looks competitive in the $\rho=1$ settings only, where clusters are designed to be Gaussian, consistently with the specified model.  
The performance of model-based clustering improve and get competitive when the data dimensionality increases, thus confirming the increase of cluster separation discussed above. 

Results from using a larger sample size ($N=3000$) are available upon request and have not been reported as essentially the same as the ones obtained with $N=1000.$ We believe that in high dimensional spaces as the ones here considered, to produce a remarkable improvement of the results, the sample size should increase to an unfeasible extent for simulation purposes.

\section{Application}
\subsection{Activity Tracking}
As first real data application, we consider a dataset describing a number of daily and sports activities measurements,  detected by 5 sensors positioned on the torso, the wrists and the sides of the knees of 8 subjects at the frequency of 25 measurements per second for 300 seconds. For each sensor location, nine variables have been recorded: the x,y,z axes acceleration, the x,y,z axes rate of turn, and the x,y,z axes Earth's magnetic field. Data are publicly available\footnote{\url{https://archive.ics.uci.edu/ml/datasets/Daily+and+Sports+Activities}} and have been extensively described by \citet{altun2010comparative, altun2010human, barshan2014recognizing}. 

For the sake of illustration, we restrict the analysis on the 3 features detected by the accelerometer of the 5 sensors in 3 different activities performed by one subject only. 
The selected activities - sitting, exercising on a cross trainer and cycling on an exercise bike - are characterized by a variety of different degree of muscular activation and force produced. Each activity has been split in 150 sub-activities of 2 seconds, hence described by 50 measurements per variable. 
The resulting data set is then formed by 450 observations of dimension $P= 15$ 
and $T=50$, grouped in three classes of activities.  

\begin{figure*}[t]
    \centering
         \begin{minipage}{\linewidth}
        \includegraphics[width =0.32 \linewidth]{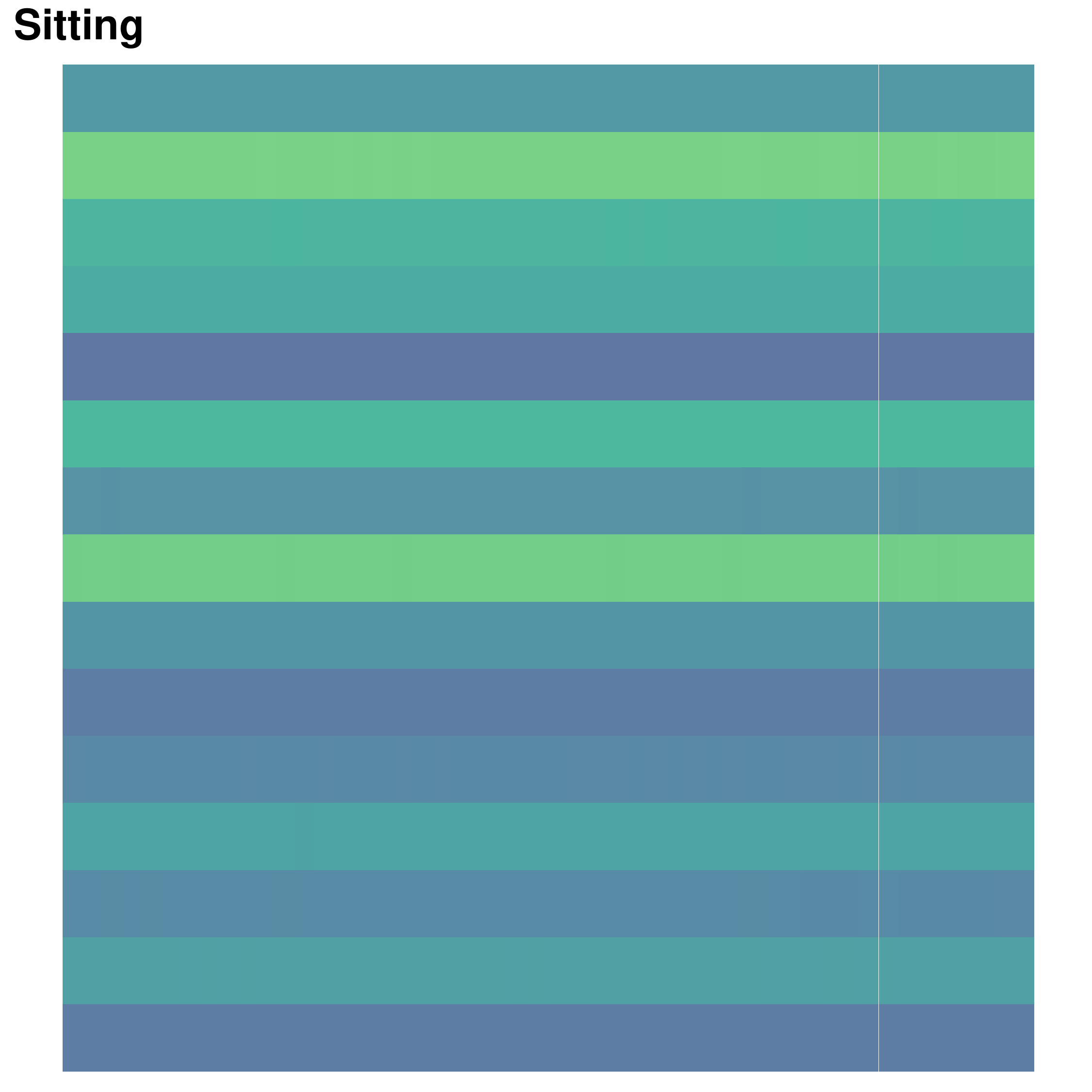}
        \includegraphics[width =0.32 \linewidth]{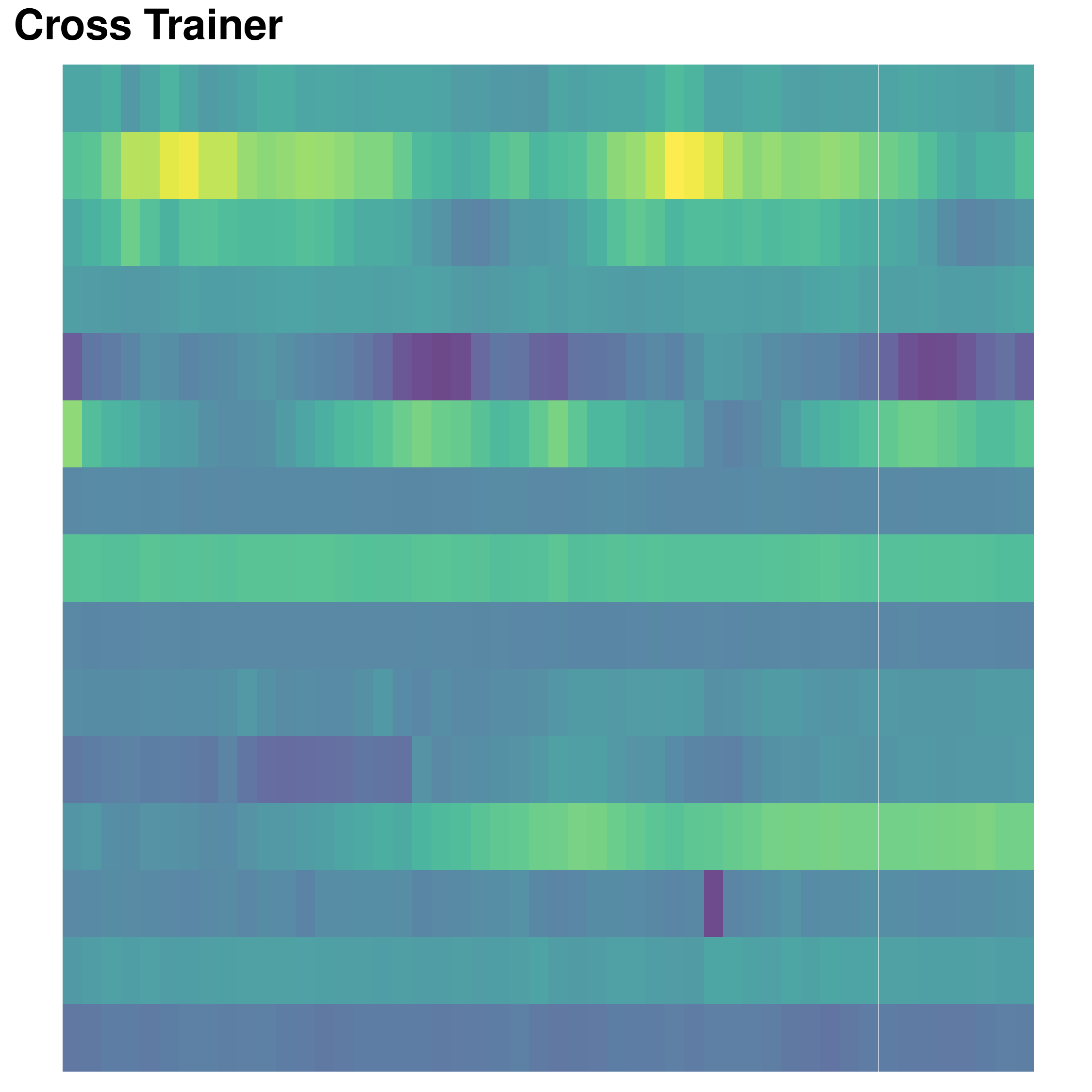}
        \includegraphics[width =0.32 \linewidth]{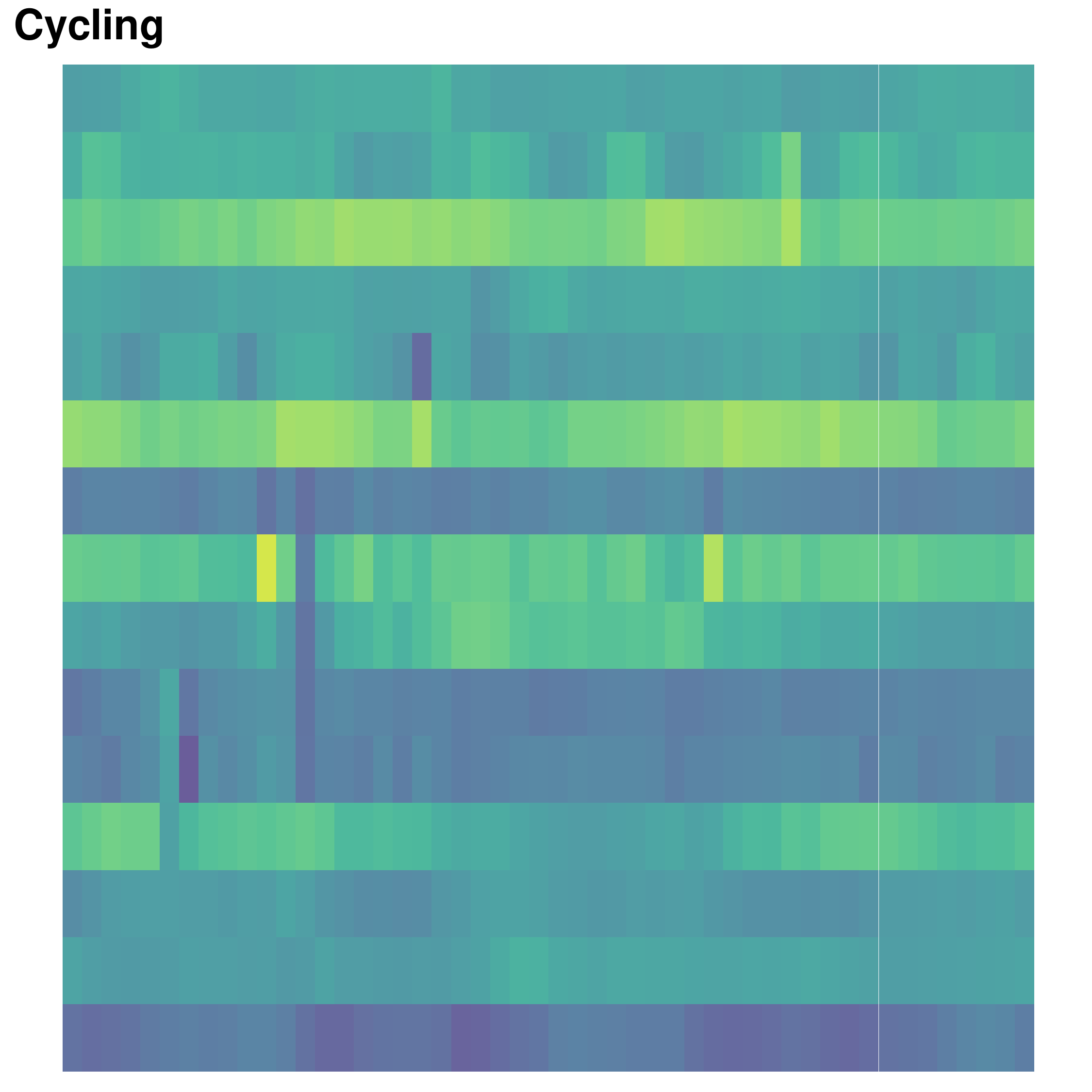}
    \end{minipage}
    \caption{Graphical representation of one observation from each of the three cluster in the activity tracking dataset. Cf. Fig. \ref{fig:centers}}
\label{fig:track}
\end{figure*}

Figure \ref{fig:track} illustrates an example of individual observation for each of the three activity. The goal of the analysis is to identify the measurements pertaining to the same activity. 
After standardizing the data, we run modal clustering based on the a $k$-NN balloon estimator, with $k= 5\sqrt N$, consistently with the indications drawn from the simulations. For comparison, we also consider the partitions detected by model-based clustering built on a mixture of Normal matrix-variate distributions and by $K$-means. The number of clusters has been selected to maximize the BIC and, respectively, the Silhouette score.  

Results, reported in Table \ref{tab:track}, show a general accuracy of all the considered methods at disclosing differences among the activities. However, while modal clustering correctly identify three groups, with just a very small amount of misclassified observations, the two competitors tend to oversegment the data, so that the actual clusters are in fact partitioned into a number smaller subgroups.  

\begin{table*}[htbp]
\caption{\emph{Tracker activity} data: comparison among partitions detected by the three clustering methods.}\label{tab:track}
\begin{center}
\begin{tabular}{l}
\\
Sitting \\
Cross Trainer\\  
  Cycling \\ 
\end{tabular}\hspace{9pt}
\begin{tabular}{rrr}
\multicolumn{3}{c}{Modal clustering}\\
  \hline
  1 & 2 & 3 \\ 
  \hline
 150 &   0 &   0 \\ 
   4 & 146 &   0 \\ 
  9 &   0 & 141 \\ 
   \hline
 \multicolumn{3}{l}{FM = 0.942}\\
 \end{tabular}\hspace{9pt}
    \begin{tabular}{rrrrrrr}
\multicolumn{7}{c}{$K$-means}\\
 \hline
  1 & 2 & 3 & 4 & 5 & 6 & 7 \\ 
  \hline
   0 &   0 &   0 &  58 &  64 &   0 &  28 \\ 
   0 &   0 &  83 &   0 &   0 &  67 &   0 \\ 
  67 &  83 &   0 &   0 &   0 &   0 &   0 \\ 
   \hline
 \multicolumn{7}{l}{FM = 0.624}\\
\end{tabular}\hspace{10pt}\hspace{10pt}
\begin{tabular}{l}
\\
Sitting \\
Cross Trainer\\  
  Cycling \\ 
\end{tabular}\hspace{9pt}
        \begin{tabular}{lrrrrrrr}
\multicolumn{7}{c}{Mixture of matrix variate Gaussians}\\        \hline
  1 & 2 & 3 & 4 & 5 & 6 & 7 \\ 
  \hline
150 &   0 &   0 &   0 &   0 &   0 &   0 \\ 
  0 &  31 &  44 &  41 &  34 &   0 &   0 \\ 
  0 &   0 &   0 &   0 &   0 &  24 & 126 \\ 
   \hline
 \multicolumn{7}{l}{FM = 0.812}\\
\end{tabular}
\end{center}
\end{table*}

\subsection{COVID-19 outgrowth across countries}

At the time of writing this paper, the whole world has been severely harmed by the COVID-19 virus, a pandemic globally causing the largest social and economic disruption since the last century. To reduce the spreading of the virus, most of countries have implemented measures of quarantine and social distancing practices, canceled or postponed most of sport, religious, political, and cultural events, interrupted business and educational activities. 

Being the virus still in action, and the overall situation still evolving, drawing general conclusions on its impact is currently not possible, also due to different information which the countries have gathered and relayed about it. It is anyway clear that the spreading and the evolution of the pandemic, as well as its impact in relation to the adopted control measures, have not been the same all over the world. With this respect, the goal of this application is to evaluate differences and similarities       
among the countries. 

The data we consider have been collected by the Oxford COVID-19 Government Response Tracker \citep[OxCGRT, ][]{covid_data} and refer to daily observations 
of the number of confirmed cases of COVID-19 in each country, the number of confirmed deaths, along with several indicators reflecting the level of government action on health policies, economic support, strictness of lockdown policies. Considered that many of these indicators are highly correlated but not always available for all the countries, our analysis accounts for just one of them, namely the Stringency index, which ranges from 0 to 100 and summarizes the government response measures to the pandemic in terms of 
schools and work spaces closing, cancellation of public events and gatherings, ``shelter-in-place'' orders, movement restrictions and the presence of informative and awareness campaigns.  

The resulting data set has been integrated with some further variables, intended to provide a rough indication about the economy and the demography of the countries. Specifically,    
the annual GDP based on purchasing power parity of each country is considered\footnote{\url{https://data.worldbank.org/indicator/NY.GDP.MKTP.CD}}, as well as the population size and the age distribution\footnote{\url{https://population.un.org/wpp/Download/Standard/Interpolated}}, grouped in the three classes of population younger than 25, from 25 to 65, and over 65 years old.

\begin{figure*}[t]
\centering
      \begin{minipage}{0.3\linewidth}
            \includegraphics[width = \linewidth]{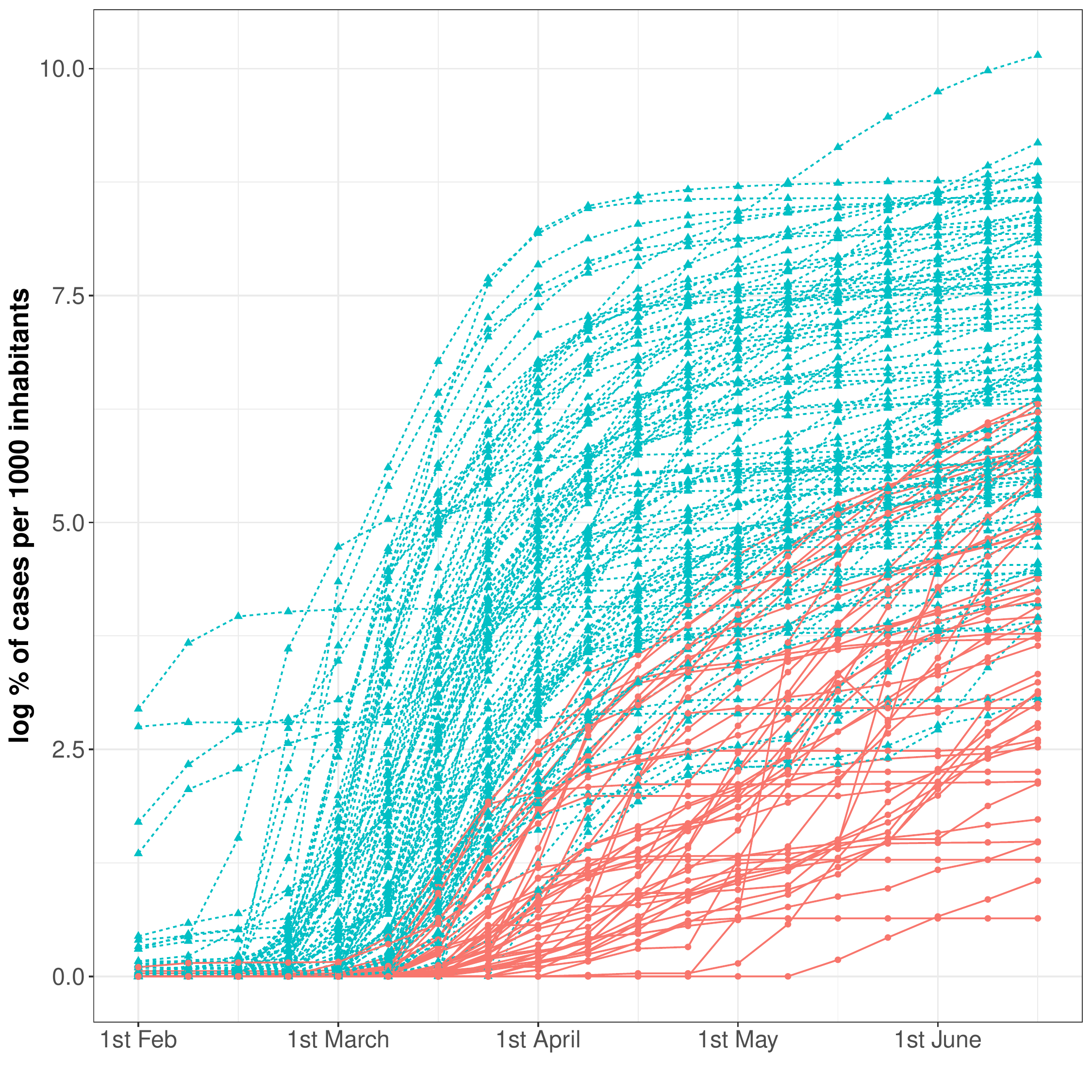}
      \end{minipage}%
        \begin{minipage}{0.3\linewidth}
            \includegraphics[width = \linewidth]{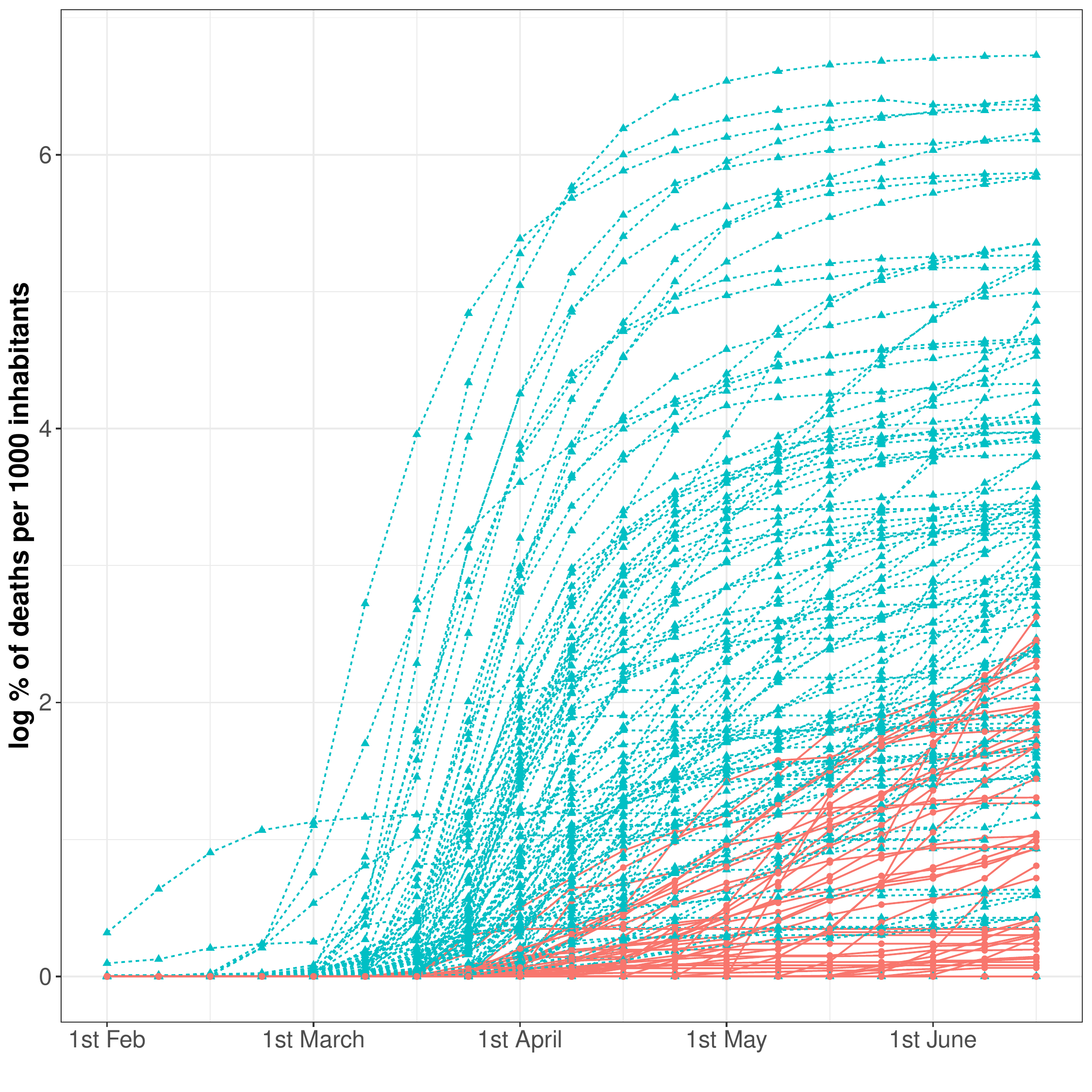}
          \end{minipage}
          \begin{minipage}{0.3\linewidth}
            \includegraphics[width = \linewidth]{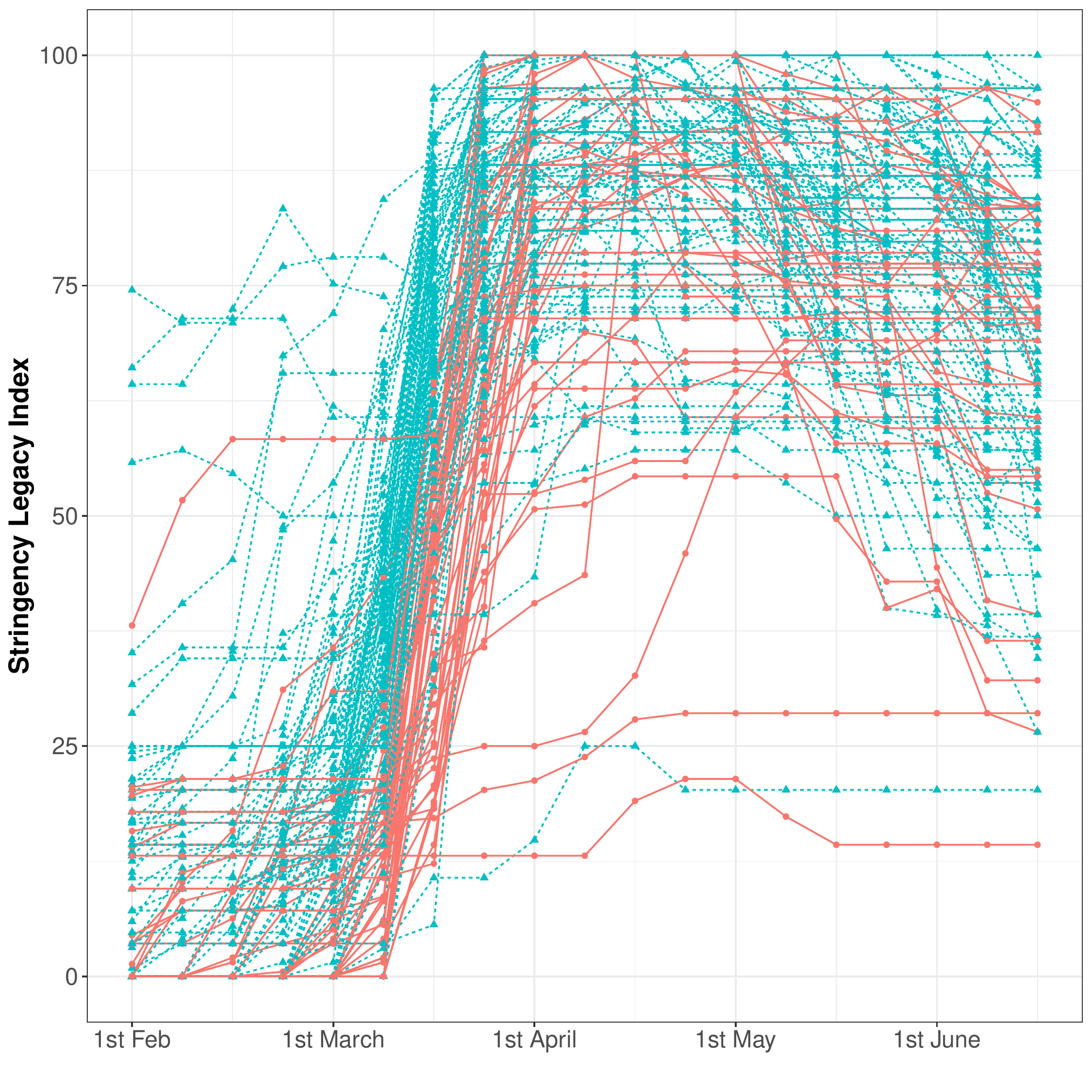}
      \end{minipage}%
    \\
      \begin{minipage}{0.3\linewidth}
            \includegraphics[width = \linewidth]{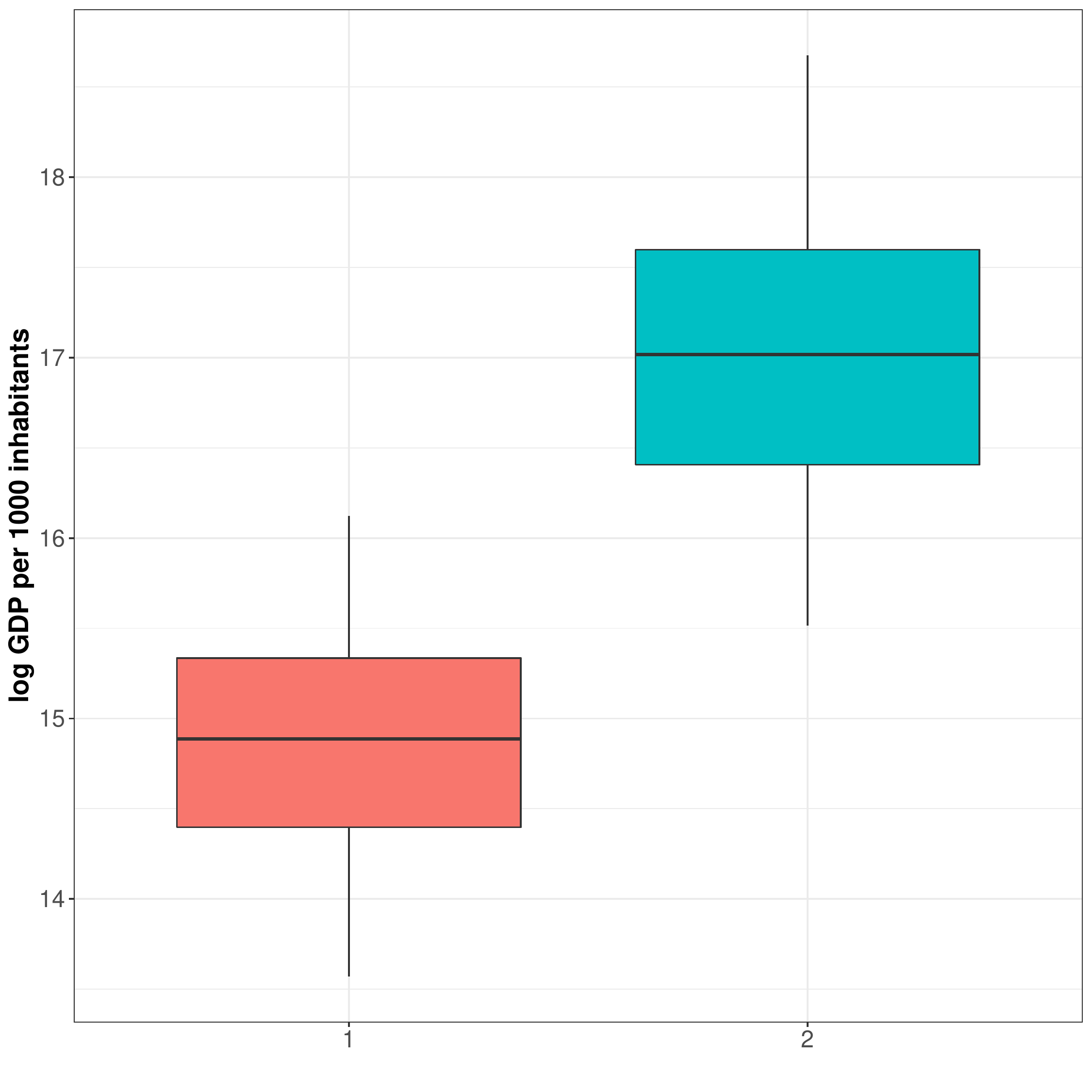}
      \end{minipage}%
        \begin{minipage}{0.3\linewidth}
            \includegraphics[width = \linewidth]{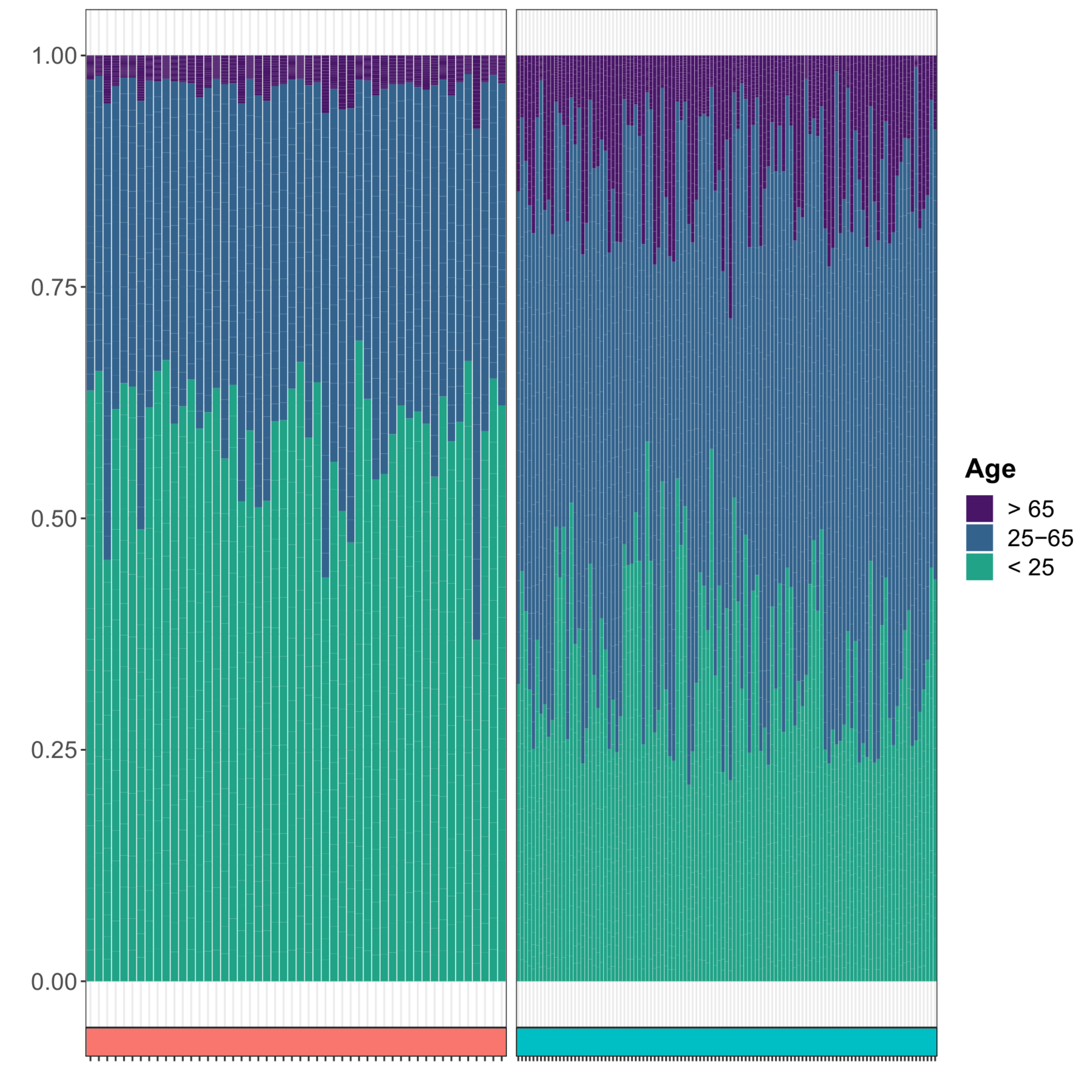}
          \end{minipage}
            \begin{minipage}{0.3\linewidth}
            \includegraphics[width = \linewidth, height = 0.7 \linewidth]{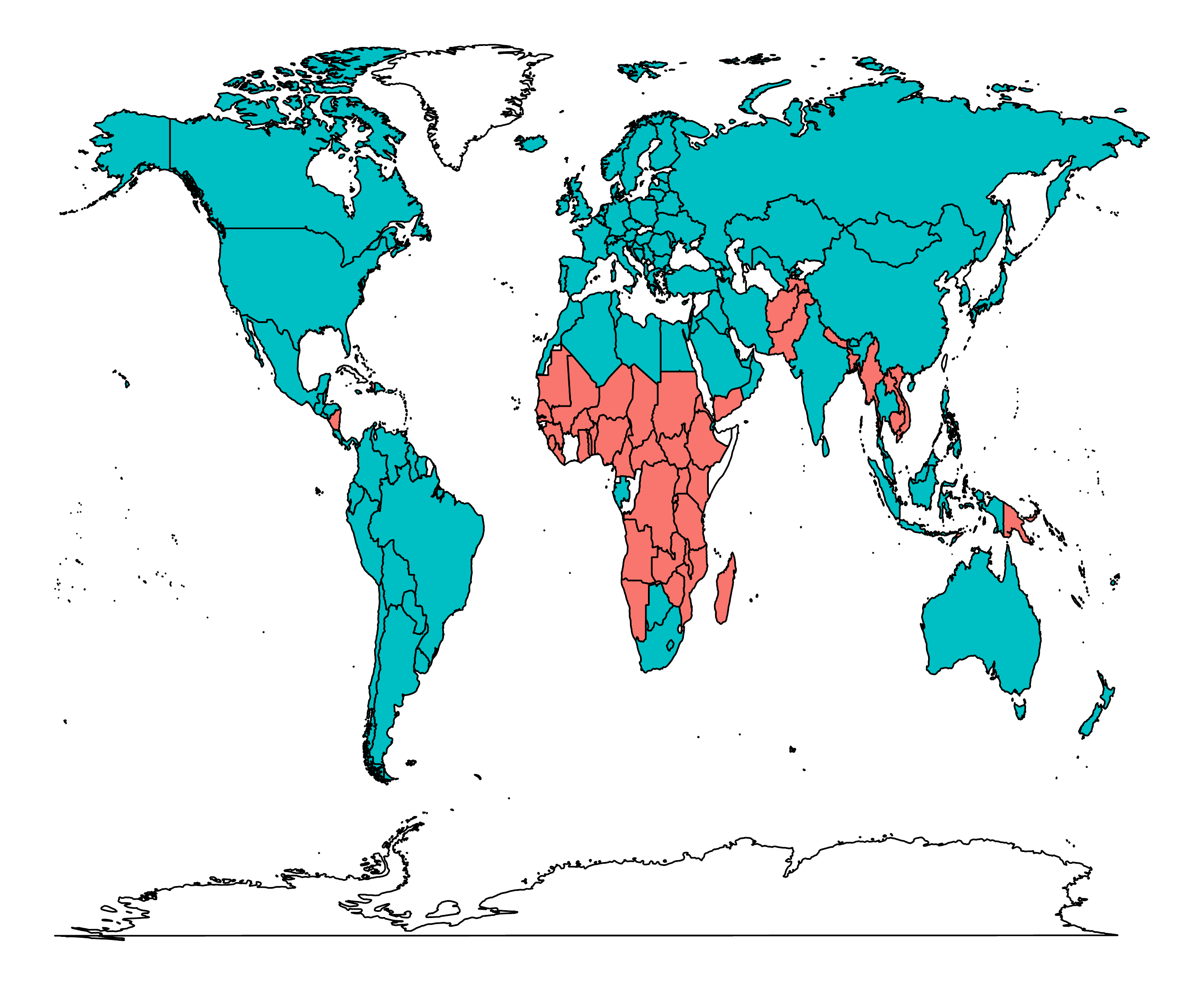}
      \end{minipage}
    \caption{Identified clusters for the considered variables in the COVID-19 study. The top panels shows the groups difference in confirmed cases, confirmed deaths, and stringency index. The bottom left panel shows the difference in GDP. The bottom middle panel shows the age distribution in the two groups. The bottom right panel shows the geographical distribution of the two identified groups.}
\label{fig:covid}
\end{figure*}

Data have been pre-processed as follows: the logarithm of the number of confirmed cases and deaths per 1000 inhabitants and of the GDP per 1000 inhabitants have been evaluated, along with the percentage of population for each of the three age classes. 
The first two variables and the Stringency index, observed on a daily basis, have been averaged to get a weekly frequency, ranging from February, 1 to June, 15. 
The final individual observation is a matrix with dimension $P = 7$ and $T = 19$. Since GDP and age distribution refer to a yearly basis, their value has been kept constant over the 19 considered weeks. All the variables have been afterwards standardized.  
A few countries have been removed from the analysis, due to the presence of missing values, thus resulting in a final sample of size $N=161$ countries. 

Similarly to the Activity Tracking example, modal clustering has been run based on a $k$-nearest neighbors balloon estimator, with $k=5\sqrt{N}.$ 
Results, illustrated in Figure \ref{fig:covid}, show an interesting pattern emerging from the data, with two clusters of countries having rather distinct characteristics. 
The largest group gathers all the countries over Europe, almost the entire America and Oceania, and many Asiatic countries, while the other group covers most of Africa and a few countries from Asia. The latter cluster is the one which the pandemic has harmed less severely in terms of both cases and deaths. While the answer of the governments, as measured by the stringency index, has not been in general weaker than in the countries assigned to the other cluster, the intervention in these countries has been in most of cases delayed, coherently with a lower perceived risk due to the limited virus spreading. Compared with the largest group, these countries have a demographic structure characterized by a larger proportion of young people, and a lower proportion of older people. This is consistent with the known behaviour of the COVID-19 virus, distressing especially older people. This apparently counterintuitive result, which labels the undeveloped countries as the less impacted by the pandemic,  has in fact rather sound motivations. On one hand, the general economic and health conditions of the undeveloped countries have likely prevented accurate testing and tracking policies, so that we shall live with a limited reliability of the data. On the other hand, the social and demographic characteristics of the undeveloped countries have likely contribute to weaken the spreading of the pandemic, due to the generally young age, a prevalent family care of older people, and a limited mobility to and from outside the country.  

It is worth noting that we ran a similar analysis also on the subset of European countries, for which further information is available (e.g. number of hospital beds per 1000 inhabitants and life expectation). European countries, taken on their own, split in two clusters, essentially formed by eastern and central Europe, and whose interpretation does not substantially depart from the one given for the whole word.

\section{Discussion}
Due to its unsupervised nature, clustering is a difficult task. The lack of an undisguised ground truth to pursue motivates a large use of visual inspection tools to get a sense of possible patterns in the data. However, high dimensionality may prevent graphical exploration to be actually fruitful, since only incomplete descriptions of the data are possible.  
Most of clustering methods are severely challenged in this framework. 
Distance-based methods, for instance, rely on the use of heuristic criteria for determining the number of clusters; on the other hand, model-based clustering requires unverifiable assumptions on the cluster shape. The scarce reliability of visual inspection tools turns then out to be rather limiting when using such approaches. 

Due to a reference cluster concept not constrained to any specific shape and to a determination of the number of groups as an integral part of the estimation procedure, modal clustering can be in principle applied even when an informative visual exploration of the data is prevented, as it may occur  with matrix-variate data. In this work we have discussed how 
this approach can be extended to three way data structures, and faced the problem both with respect to the issue of density estimation and the one of mode detection. 

Building on the use of nonparametric tools, the great challenge, apparently discouraging from the use of modal clustering in the considered setting, is its known disruption in high dimensional spaces, as matrix-valued data turn out to be intrinsically. Simple rules of thumb to select the smoothing amount in density estimation have proved moderate accuracy in nontrivial settings, at least with limited sample sizes as the one explored in this work. Indeed, the arising of small spurious clusters often hampers the application of nonparametric methods for density estimation to matrix data.

In fact, numerical explorations performed in this work have shown that the gross clustering structure is usually identified. Since cluster separation tends to increase with the data dimensionality, the situation is not that critical as it might in principle appear. While often the problem of spurious clusters cannot be straightened out completely, modal clustering has proven extraordinarily effective even the matrix overall dimension is in the order of several hundreds and exceeds the sample size. 
Adaptive tools which account for the local characteristics of the data have proven to be quite effective in this context. Defining the amount of (local) smoothing, here intended as the proportion of sample neighbors to account for, is still an  open problem. 
In our exploration we have considered simple heuristic criteria, highlighting that a large amount of smoothing is usually advisable, especially when the matrix dimension is large. However, defining more rigorous criteria targeted to the specific problem would be desirable and is left for future work.



\appendix
\section*{Appendix A: Proof of Proposition 1}\label{proof1}

    To establish the expression for the AMISE in Proposition 1, we start from the standard decomposition \eqref{eq:mise} 
    and consider its characterization in terms of asymptotic IV and ISB. \\
	Let us start by analyzing the asymptotic behaviour of the bias term. With a change of variables, the expected value of $\hat{f}(X; h)$ may be expressed as
	\begin{align}
	\mathbb{E}(\hat{f}(X; h)) &= \int_{\mathbb{R}^{P,T}} h^{-P\cdot T}K(h^{-1}(X - Y))f(Y)dY \nonumber\\
	&= \int_{\mathbb{R}^{P,T}} K(Z)f(X - hZ)dZ .
	\label{eq:exp_value}
  	\end{align}
The term $f(X - hZ)$ may be approximated with a Taylor expansion around $X$, thus obtaining
	\begin{equation*}
	f(X - hZ) = f(X) - h \, \text{tr}(\nabla f(X)^{\top} Z) + \frac{1}{2}h^2\text{tr}(\textsf{H}f(X)^{\top}  Z\otimes Z) + o(h^{2}),
	\end{equation*}
	where $\textsf{H}$ denotes the Hessian matrix of dimensions $PT \times PT$. Using the fact that $\int_{\mathbb{R}^{P,T}}K(X)dX = 1$ and that $\int_{\mathbb{R}^{P,T}}X K(X)dX = 0$, and plugging the Taylor expansion into Equation \eqref{eq:exp_value}, we get
	\begin{align*}
	\mathbb{E}(\hat{f}(X; h)) &= f(X) + \frac{1}{2}h^{2}\text{tr}(\textsf{H}f(X)^{\top}m_{2}(K)\mathbb{I}_{P\cdot T}) + o(h^2) \\
	&= f(X) + \frac{1}{2}h^{2}m_{2}(K)\text{tr}(\textsf{H}f(X)) + o(h^{2}) \\
	&= f(X) + \frac{1}{2}h^{2}m_{2}(K)\Delta f + o(h^{2}) .
	\end{align*}
Hence the approximated squared bias is 
\begin{equation*}
    	[\mathbb{E}(\hat{f}(X; h))  - f(X)]^2 = \frac{1}{4}h^{4}m_{2}^2(K)\Delta f^2 + o(h^{4}).
\end{equation*}
By integrating with respect to $X$ we obtain
	\begin{equation}
	\text{ISB}(\hat{f}(\cdot; h)) = \frac{1}{4}h^{4}m_{2}(K)^{2}R(\Delta f) + o(h^{4}).
	\label{eq:ISB}
	\end{equation}
The variance of $\hat{f}$ is given by
    \begin{align}
        \label{eq:var}
        \text{Var}(\hat{f}(X; h)) &= N^{-1}\int_{\mathbb{R}^{P,T}}h^{-2P\cdot T}K(h^{-1}(X - Y))^{2}f(Y)dY -  \nonumber\\ 
            &\qquad  N^{-1}\left(\int_{\mathbb{R}^{P,T}}h^{-P\cdot T}K(h^{-1}(X - Y))f(Y)dY\right)^{2} .
    \end{align}
    Starting from the first term in \eqref{eq:var}, and integrating it with respect to $X$, we obtain
	\begin{align}
	& N^{-1}\int_{\mathbb{R}^{P,T}}\int_{\mathbb{R}^{P,T}}h^{-2P\cdot T}K(h^{-1}(X - Y))^{2}f(Y)dYdX \nonumber \\
	&\qquad = N^{-1}h^{-(P \cdot T)}\int_{\mathbb{R}^{P,T}}\int_{\mathbb{R}^{P,T}}K(Z)^{2}f(X - hZ)dZdX \nonumber\\
	&\qquad = N^{-1}h^{-(P \cdot T)}R(K),
	\label{eq:IV_int}
	\end{align}
	where the first equality follows from the change of variable $Z = h(X - Y)$ and the second one from Fubini's theorem. For the second term in equation \eqref{eq:var} we can take advantage of the previous calculations for $\mathbb{E}(\hat{f}(X; h))$ to obtain 
	\begin{align*}
	&N^{-1}\int_{\mathbb{R}^{P,T}}\int_{\mathbb{R}^{P,T}}h^{-2P\cdot T}K(h^{-1}(X - Y))^{2}f(Y)dYdX \\
	& \qquad = N^{-1}R(f) + o(N^{-1}).
	\end{align*}
Given the assumption (iii), and in view of equation \eqref{eq:IV_int}, it follows that this second term in the IV is of a smaller order than the first one. Therefore, 
	\begin{equation}
	\text{IV}(\hat{f}(\cdot; h)) = N^{-1}h^{-(P \cdot T)}R(K) + o(N^{-1}h^{-(P \cdot T)}).
	\label{eq:IV}
	\end{equation}
	Combining Equations \eqref{eq:ISB} and \eqref{eq:IV}, it follows that an asymptotic approximation to the MISE can be written as
	\begin{equation*}
	\text{AMISE}(\hat{f}(\cdot; h)) = N^{-1}h^{-(P \cdot T)}R(K) + \frac{1}{4}h^{4}m_{2}(K)^{2}R(\Delta f) .
	\end{equation*}
The optimal bandwidth \eqref{eq:h_opt} is derived via minimization of the AMISE, by identifying the root of 
%
\begin{equation*}
    \frac{\partial}{\partial h}\text{AMISE}(\hat{f}(\cdot; h)) = -\frac{(P\cdot T)R(K)}{N h^{-(P\cdot T) - 1}} + h^{3}m_{2}(K)^{2}R(\Delta f) = 0.
\end{equation*}

\section*{Appendix B: Proof of Proposition 2}

Since the maxima of a function $f$ satisfies $\nabla f = 0,$ a standard formulation of a gradient ascent algorithm on its estimate s the following:
 \begin{equation}\label{eq:ga}
    Y^{(s+1)} = Y^{(s)} + \alpha \nabla \hat{f}(Y^{(s)}).
\end{equation}
For the specific case \eqref{eq:kde}, simple differentiation rules of matrix-variate function lead to \begin{equation}
\label{eq:grad}
    \nabla \hat{f}(X; h) = \frac{1}{Nh^{P\cdot T}}\sum_{n = 1}^{N} \nabla K\left(h^{-1}(X_{n} - X)\right) .
\end{equation}
The use of a spherically symmetric kernel $K$ allows recasting to the simpler use of a function of real variable $\kappa: \mathbb{R}_+ \mapsto \mathbb{R}$, known as \textit{profile} of $K$, via the representation
\begin{equation*}
    K(X) = \frac{1}{2}\kappa(||X||_{F}^{2}) = \frac{1}{2}\kappa(\text{tr}(X^{\top}X)) , 
\end{equation*}
where $||\cdot||_{F}$ is the Frobenius norm. Note that the first equality highlights the similar structure of the matrix-variate kernel to the standard multivariate kernel, where the Euclidean norm replaces the Frobenius norm. Hence, the \eqref{eq:grad} turns into the following: 

\begin{align}
\label{eq:grad2}
    \nabla \hat{f}(X; h) 
    &= \frac{1}{N}\frac{1}{2 h^{P\cdot T}} \sum_{n = 1}^{N} \nabla \kappa\left(h^{-2}\text{tr}((X_{n} - X)^\top(X_{n} - X))\right) \nonumber\\
    &= -\frac{1}{N}\frac{1}{h^{P\cdot T + 2}} \sum_{n = 1}^{N} \kappa'\left(h^{-2}\text{tr}((X_{n} - X)^\top(X_{n} - X))\right) (X_{n} - X).
\end{align}
Replacing the \eqref{eq:grad2} in the $\nabla \hat{f}$ term of \eqref{eq:ga}, we obtain: \\
\resizebox{.98\linewidth}{!}{
  \begin{minipage}{\linewidth}
\begin{align}
\label{eq:ga_wm}
    Y^{(s+1)} &= Y^{(s)} - \alpha  \frac{1}{N h^{P\cdot T + 2}} \sum_{n = 1}^{N} \kappa'\left(h^{-2}\text{tr}((X_{n} - X)^\top(X_{n} - Y^{(s)}))\right) (X_{n} - Y^{(s)}) \nonumber\\
    &= Y^{(s)} - \alpha \left[ \frac{1}{N}\frac{1}{h^{P\cdot T + 2}} \sum_{n = 1}^{N} \kappa'\left(h^{-2}\text{tr}((X_{n} - X)^\top(X_{n} - Y^{(s)}))\right) X_{n} + \right. \nonumber\\
    &\quad \left. \frac{1}{N}\frac{1}{h^{P\cdot T + 2}} \sum_{n = 1}^{N} \kappa'\left(h^{-2}\text{tr}((X_{n} - X)^\top(X_{n} - Y^{(s)}))\right) Y^{(s)}\right],  
\end{align}
 \end{minipage}
}\\

\noindent and setting an adaptive step size 
\begin{equation*}
    \alpha = \alpha_{s} = \left[-\frac{1}{N}\frac{1}{h^{P\cdot T + 2}}  \sum_{n = 1}^{N} \kappa'\left(h^{-2}\text{tr}((X_{n} - Y^{(s)})^\top(X_{n} - Y^{(s)}))\right)\right]^{-1}.
\end{equation*} 
we obtain the thesis. 

\bigskip

\bibliographystyle{elsarticle-harv}      
\bibliography{bib}   

\end{document}